\documentclass[aps,prc,twocolumn,floatfix,superscriptaddress]{revtex4}
\usepackage{epsfig}
\usepackage{amsmath}
\usepackage{color}

\usepackage{multirow}
\usepackage{graphicx}

\definecolor{blue}{rgb}{0.05, 0.05, 0.5}

\begin{document}

\title{ Initial Temperature and Extent of Chemical Equilibration of Partons in Relativistic Collision of  Heavy Nuclei}
\author{Dinesh K. Srivastava}
\email{dinesh@vecc.gov.in}
\affiliation{Variable Energy Cyclotron Centre, HBNI, 1/AF, Bidhan Nagar, Kolkata 700064, India}
\author{Rupa Chatterjee}
\email{rupa@vecc.gov.in}
\affiliation{Variable Energy Cyclotron Centre, HBNI, 1/AF, Bidhan Nagar, Kolkata 700064, India}
\author{Munshi G. Mustafa}
\email{munshigolam.mustafa@saha.ac.in}
\affiliation{Saha Institute of Nuclear Physics, HBNI, 1/AF Bidhan Nagar, Kolkata 700064, India}

\begin{abstract}
We emphasize that a knowledge of energy and entropy densities of  quark gluon plasma - a thermalized 
de-confined matter, formed in relativistic heavy ion collisions fixes the formation temperature and the product of gluon fugacity and formation time uniquely, {\em provided} we know the relative fugacities of quarks and gluons. This also provides that a smaller formation time would imply larger fugacities for partons. Next we explore the limits of chemical equilibration of partons during the initial stages in relativistic collision of heavy nuclei. The experimentally measured rapidity densities  of transverse energy and charged particle multiplicity at RHIC and LHC energies  are used to estimate the energy and number densities with the assumption of formation of a thermally equilibrated quark gluon plasma which may be chemically equilibrated to the same or differing extents for quarks and gluons. The estimates are found to be very sensitive to the correction factor used for the Bj\"{o}rken energy density for identifying it with the initial energy density. The extent of chemical equilibration near the end of the QGP phase is inferred by solving master equations by including the processes $gg \leftrightarrow ggg$ and $gg \leftrightarrow q\overline{q}$ along with expansion and cooling of the plasma. The possible consequences for invariant mass distribution of intermediate mass dileptons radiated from the plasma are discussed which could distinguish between different scenarios.
\end{abstract}


\maketitle

\maketitle
\section{Introduction} 
The discovery and characterization of the properties of quark gluon plasma, the strongly interacting deconfined matter, remains one of the best orchestrated international efforts in modern nuclear physics.  The Relativistic Heavy Ion Collider (RHIC) and the Large Hadron Collider (LHC) studying the collisions of heavy nuclei at relativistic energies continue to generate a wealth of data which is being analyzed to provide valuable information about the nature of the ephemeral matter thus created. It is now believed that a matter having a temperature of a few hundred MeV and an energy density of several GeV/fm$^3$ is being created in these collisions and that it is partonic in nature. Several  signatures, envisaged and not necessarily envisaged earlier, {\it viz.}, radiation of photons and dileptons, suppression of hadrons having large transverse momenta due to jet-quenching, suppression (and regeneration) of heavy quarkonia, elliptic (and higher order) flow etc., have been confirmed~\cite{hwa_wang}.

The early theoretical considerations using pQCD to study initial conditions, supplemented with concepts of self-screening~\cite{sspc} and parton saturation~\cite{kari} or cascading partons~\cite{klaus,bms,carsten} have provided valuable insights while the developments in concepts of colour gluon condensate model\cite{raju} have reached a level of sophistication, where the results are able to provide detailed description of
the experimental data, most notably the particle spectra and their azimuthal anisotropy.

Our goal in the present work is rather modest. We recall that at least the description of the electromagnetic radiations from the plasma requires a detailed knowledge of the constituents of the plasma in terms of their quark and gluon contents. In view of this, the present work explores the limits of the chemical equilibration of the partons in the initial stages expected on the basis of the rapidity densities of the particle multiplicity and transverse energy  measured in these collisions. The transverse energy of the collision is often used to estimate the initial energy density using the so-called Bj\"{o}rken energy estimate~\cite{bjorken}, which can be considered as its lower limit. We introduce a correction factor $f_{\text{Bj}}$ to denote this and explore the dependence of our results on this. 

\section{Formulation} 
We realize that at relativistic energies, a large number of low-$x$ partons from the colliding nuclei have sufficiently large linear momenta. The vehement collisions of  the partons and their subsequent multiplications lead to  a rapid (local) thermalization  at an initial time $\tau_0$. This could be of the order of a fraction of a fm/$c$ at LHC energies and larger at the smaller energies used, say, for the beam energy scan program at RHIC. 

We further assume that the evolution of the system at later times can be described by relativistic hydrodynamics. In these initial studies we neglect the viscosity of the system for simplicity. We assume that the partons, limited to light quarks and gluons, may or may not be in a state of chemical equilibrium. The chemical equilibration, if not already attained, proceeds via the processes $gg \leftrightarrow ggg$  and $ gg \leftrightarrow q\overline{q}$, for which we set up the necessary master equations~\cite{biro, munshi,duncan, wang, charm}. The elastic scatterings are assumed to maintain the kinetic equilibrium during the expansion and cooling. 

We also neglect the transverse expansion of the plasma in these exploratory calculations. However it can be easily incorporated~\cite{munshi, duncan, munshi_str}. 
We shall closely follow the notations etc. from Ref.~\cite{biro} in the following, except for the rate $R_3$ (see later) for which  a more recent expression derived in Ref.~\cite{wang} has been used.

Assuming the plasma thus formed to be an ideal fluid its expansion can be described by the equation for conservation of energy and momentum:
\begin{equation}
\partial_\mu T^{\mu \nu}=0 \; , \qquad
 T^{\mu \nu}=(\epsilon+P) u^\mu u^\nu + P g^{\mu \nu} \, ,
\label{hydro}
\end{equation}
where $\epsilon$ is the energy density and $P$ is the pressure measured in the frame comoving with the fluid and $u^\mu$ is the four-velocity vector of the fluid with the constraint $u^2=-1$. We shall describe  a partially equilibrated plasma using scaled distribution functions for equilibrium distribution of  gluons and  quarks:
\begin{equation}
g_{i}(q,T,\lambda_i)= \lambda_{i} g^{\mathrm{eq}}_i(q,T) \, ,
 \label{befd}
\end{equation}
where
\begin{equation}
 g^{\mathrm{eq}}_i(q,T)=1/({e^{\beta{u\cdot q}}\pm 1})
\end{equation}
is the Fermi-Dirac (Bose-Einstein) distribution for quarks (gluons),
 $\beta$ is $1/T$ and $\lambda_i$ is the fugacity for parton species $i$ and
describes the extent of its deviation from chemical equilibrium. 
This accounts for the under-saturation or over-saturation of the parton phase space density. Thus for undersaturaion of the species $i$,  
$0\leq\lambda_i\leq 1$ while
$\lambda_i \geq 1$ for its over-saturation. The latter has been
 discussed by some authors
in connection with thermal models describing the ratios of particle
production~\cite{rafelski}.

We write the equation of state for a partially equilibrated plasma of mass-less
particles as
\cite{biro}
\begin{equation}
\epsilon=3P= \left [a_2 \lambda_g +  b_2 \left (\lambda_q+\lambda_{\bar q}
\right ) \right ] T^4 \, ,
\label{eos}
\end{equation}
where $a_2=8\pi^2/15$, $b_2=7\pi^2 N_f/40$ and $N_f \approx 2.5$ is
the number of dynamical quark flavours.  The density of an
equilibrating partonic system can be written as
\begin{equation}
n_g=\lambda_g \tilde{n}_g,\qquad
 n_q=\lambda_q \tilde{n}_q,
\end{equation}
where $\tilde{n}_k$ is the equilibrium density for the parton species $k$:
\begin{equation}
\tilde{n}_g=\frac{16}{\pi^2}\zeta(3) T^3=a_1 T^3,
\end{equation}
\begin{equation}
\tilde{n}_q=\frac{9}{2\pi^2}\zeta(3) N_f T^3=b_1 T^3,
\end{equation}
so that the number density of the partons reduces to
\begin{equation}
n= \left [n_g+n_q+n_{\bar{q}} \right]= \left [a_1 \lambda_g+b_1 (\lambda_q+\lambda_{\bar{q}}) \right]T^3.
\end{equation}
We neglect the net baryons in the central rapidity region and
assume that $\lambda_q=\lambda_{\bar{q}}$. This is definitely valid at central rapidities at larger
energies at RHIC and LHC and will remain a reasonable approximation, 
as long  as the number of  produced pions far exceeds the number of net baryons.
This equation of state (Eq.~\ref{eos}) provides that
the speed of sound $c_s=1/\sqrt{3}$.

Now the master equations \cite{biro} for the dominant chemical reactions
$gg \leftrightarrow ggg$ and $gg \leftrightarrow q\bar{q}$  are written as:
\begin{eqnarray}
\partial_\mu (n_g u^\mu)&=&n_g(R_{2 \rightarrow 3} -R_{3 \rightarrow 2})
                    - (n_g R_{g \rightarrow q}
                       -n_q R_{q \rightarrow g} ) \, , \nonumber\\
\partial_\mu (n_q u^\mu)&=&\partial_\mu (n_{\bar{q}} u^\mu)
                     = n_g R_{g \rightarrow q}
                       -n_q R_{q \rightarrow g},
\label{master1}
\end{eqnarray}
in an obvious notation. 

It is well known that for a purely longitudinal boost invariant expansion, 
Eq.~\ref{hydro} reduces to~\cite{bjorken}:
\begin{equation}
\frac{d\epsilon}{d\tau}+\frac{\epsilon+P}{\tau}=0,
\label{long}
\end{equation}
where $\tau$ is the proper time. This equation implies
\begin{equation}
\epsilon\, \tau^{4/3}=\,{\mathrm {const}.}
\label{epstau}
\end{equation}

The master equations governing the chemical equilibration reduce to~\cite{biro}
\begin{eqnarray}
\frac{1}{\lambda_g}\frac{d \lambda_g}{d\tau}
+\frac{3}{T}\frac{dT}{d\tau} +
\frac{1}{\tau}
 &=&
R_3 ( 1- \lambda_g ) -2 R_2 \left( 1-\frac{\lambda_q \lambda_{\bar{q}}}
{\lambda_g^2}\right) \, ,
\nonumber\\
\frac{1}{\lambda_q}\frac{d \lambda_q}{d\tau}
+\frac{3}{T}\frac{dT}{d\tau} +
\frac{1}{\tau}
 &=&
R_2 \frac{a_1}{b_1} \left(
\frac{\lambda_g}{\lambda_q}-\frac{\lambda_{\bar{q}}}{\lambda_g}\right)\, ,
\label{master_long}
\end{eqnarray}
which are solved numerically for the fugacities for any given initial condition.

We quote the rate constants $R_2$ and $R_3$ appearing in Eq.~(\ref{master1})
 for the sake of completeness. The rate constant $R_2$ is taken from the 
Ref~\cite{biro}, while $R_3$ is taken from Ref.~\cite{wang}:
\begin{eqnarray}
R_2 & \approx & 0.24 N_f \alpha_s^2 \lambda_g T \ln (1.65/\alpha_s \lambda_g),
\nonumber\\
R_3 &= &\frac{32}{3a_1} \alpha_s T \lambda_g \left ( 1 + \frac{8}{9}
a_1\alpha_s \lambda_g \right )^2 {\cal I}(\lambda_g) \ \ ,
\label{R_3}
\end{eqnarray}
where the colour Debye screening and the Landau - Pomeranchuk - Migdal effect
suppressing the induced gluon radiation have been taken into account,
explicitly and
\begin{eqnarray}
{\cal I}(\lambda_g) &=& \int_1^{{\sqrt s}\lambda_f} dx \int_0^{s/4\mu_D^2} dz
{\frac{z}{(1+z)^2}} \nonumber \\
&& \left\{
\frac {\cosh^{-1} \left ( \sqrt x \right )}
{x \sqrt{\left [ x + (1+z)x_D\right ]^2 -4xzx_D}} \right. \nonumber \\
&& \left. + \frac{1}{s\lambda_f^2}
\frac {\cosh^{-1} \left ( \sqrt x \right )}
{\sqrt{\left [ 1 + x(1+z)y_D\right ]^2 -4xzy_D}} \
\right \} ,
\end{eqnarray}
with $x_D= \mu_D^2\lambda_f^2$, $y_D=\mu_D^2/s$,
 and $\mu_D^2=4 \pi \alpha_s T^2 \lambda_g$.
We also note that $\lambda_f$ is the mean-free path for
 elastic scattering given by
\begin{equation}
\lambda_f^{-1}=\frac{9}{8}a_1 \alpha_s T \frac{1}{1+8 \pi \alpha_s \lambda_g/9}.
\end{equation}

We conclude this section by  recalling that the Bj\"{o}rken energy density 
$\epsilon_{\text{Bj}}$ is defined as
\begin{equation}
\epsilon_{\text{Bj}}(\tau_0)=\frac{1}{\pi R_T^2 \tau_0} \frac{dE_T}{dy}
\label{bjorken}
\end{equation}
where $R_T$ is the transverse dimension of the system and $dE_T/dy$ is the 
transverse energy of the produced particles in the central rapidity
region. 

This can be identified with the initial energy density $\epsilon(\tau_0)$
{\em if} the work-done against the pressure during the longitudinal expansion 
and effects of viscosity can be neglected. The transverse expansion is expected to only redistribute the available transverse energy 
among the particles. It has been argued that a proper accounting for these
increases the Bj\"{o}rken energy estimate by a factor of 2~\cite{mikolos}. In what
follows we define:
\begin{equation}
\epsilon(\tau_0)=f_\text{Bj} \epsilon_\text{Bj}(\tau_0),
\label{us}
\end{equation}
where $f_\text{Bj}\approx 2$ stands for the corrections to the Bj\"{o}rken
estimate for the initial energy density~\cite{mikolos}. We shall see later 
that our results will depend crucially on the inclusion/neglect of this correction
factor.

Proceeding along similar lines but now neglecting any change in entropy during the expansion, the initial number
density of the partons can be written as
\begin{equation}
n(\tau_0)=\frac{1}{\pi R_T^2 \tau_0} \frac{dN}{dy}
\label{density}
\end{equation}
where $N$ is the multiplicity of the particles produced. 

In what follows we use the experimentally measured pseudo-rapidity density 
of the transverse energy and the multiplicity of charged particles obtained
experimentally at RHIC~\cite{rhic} and LHC~\cite{lhc} to estimate these as
follows:
\begin{equation}
\frac{dE_T}{dy}=J(y,\eta) \frac{dE_T}{d\eta}
\end{equation}
and
\begin{equation}
\frac{dN}{dy}=\frac{3}{2} J(y,\eta) \frac{dN_{\text{ch}}}{d\eta}
\end{equation}
where $J(y,\eta)$ stands for the Jacobian for the $\eta$ to $y$ transformation
(see Ref.~\cite{rhic,lhc})
and the factor $3/2$ accounts for the neutral particles.

\section{Initial conditions, chemical evolution and radiation of dileptons}

Writing the general expressions for the energy and number densities at the initial time $\tau_0$ :

\begin{equation}
\epsilon(\tau_0)= \left [\lambda_g a_2 \, + \, 2 \lambda_q b_2 \right] T^4_0
\end{equation}
and
\begin{equation}
n(\tau_0)= \left [\lambda_g a_1 \, + \, 2 \lambda_q b_1 \right] T^3_0 \, 
\end{equation}
we get
\begin{equation}
T_0 = \left [ \frac{\epsilon(\tau_0)}{n(\tau_0)}\right ] \times  \left [ \frac{a_1 \, + \, 2 \alpha b_1}{ a_2 \, + \, 2\alpha b_2} \right] \, ,
\end{equation}
where $\alpha=\lambda_q/\lambda_g$ is the relative fugacities of quarks and gluons. We can, in principle, estimate the initial energy density $\epsilon$ in terms of the experimentally measured rapidity density of the transverse energy and the number (or equivalently entropy) density in terms of the experimentally determined rapidity density of produced particles. This then fixes the initial temperature in terms of $\alpha$. The product $\lambda_g(\tau_0)\tau_0$ can then be obtained from the rapidity density of the particles (see again later).
In the following we explore specific cases by using some estimates for $f_{\rm {Bj}}$, $\alpha$, and $\tau_0$.  

We already have the necessary mathematical frame-work in place to study the chemical evolution
of the plasma formed in relativistic collision of nuclei. As mentioned above this has been studied in several early works~\cite{biro,munshi,duncan} and extended to study equilibration of strangeness~\cite{munshi_str} and thermal production of charm~\cite{wang} using initial conditions derived from a self screened
parton cascade model~\cite{sspc} or HIJING model~\cite{hijing}.

\subsection{Thermally and chemically equilibrated quark gluon plasma}

\begin{figure}
\centerline{\includegraphics*[width=8.0 cm]{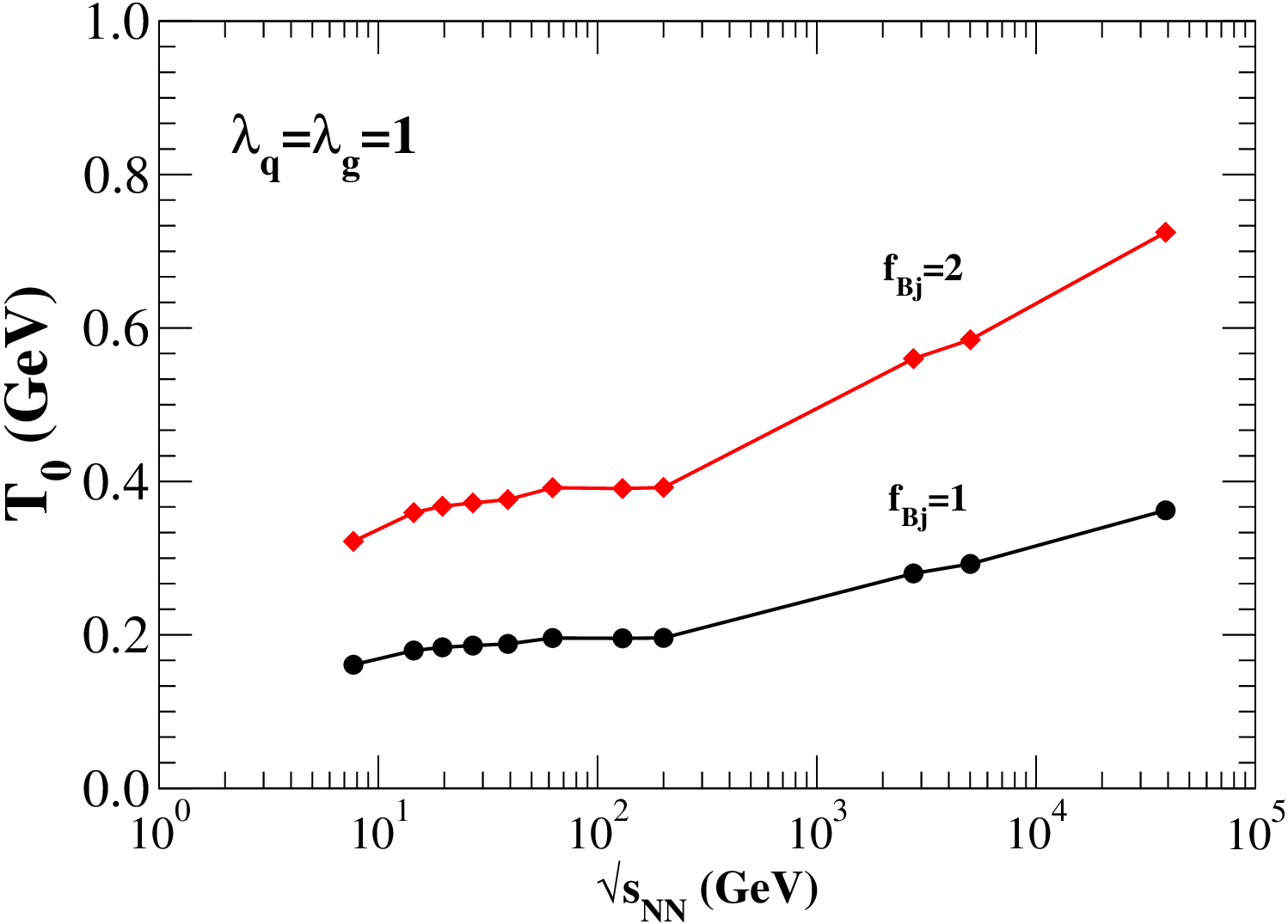}}
\centerline{\includegraphics*[width=8.0 cm]{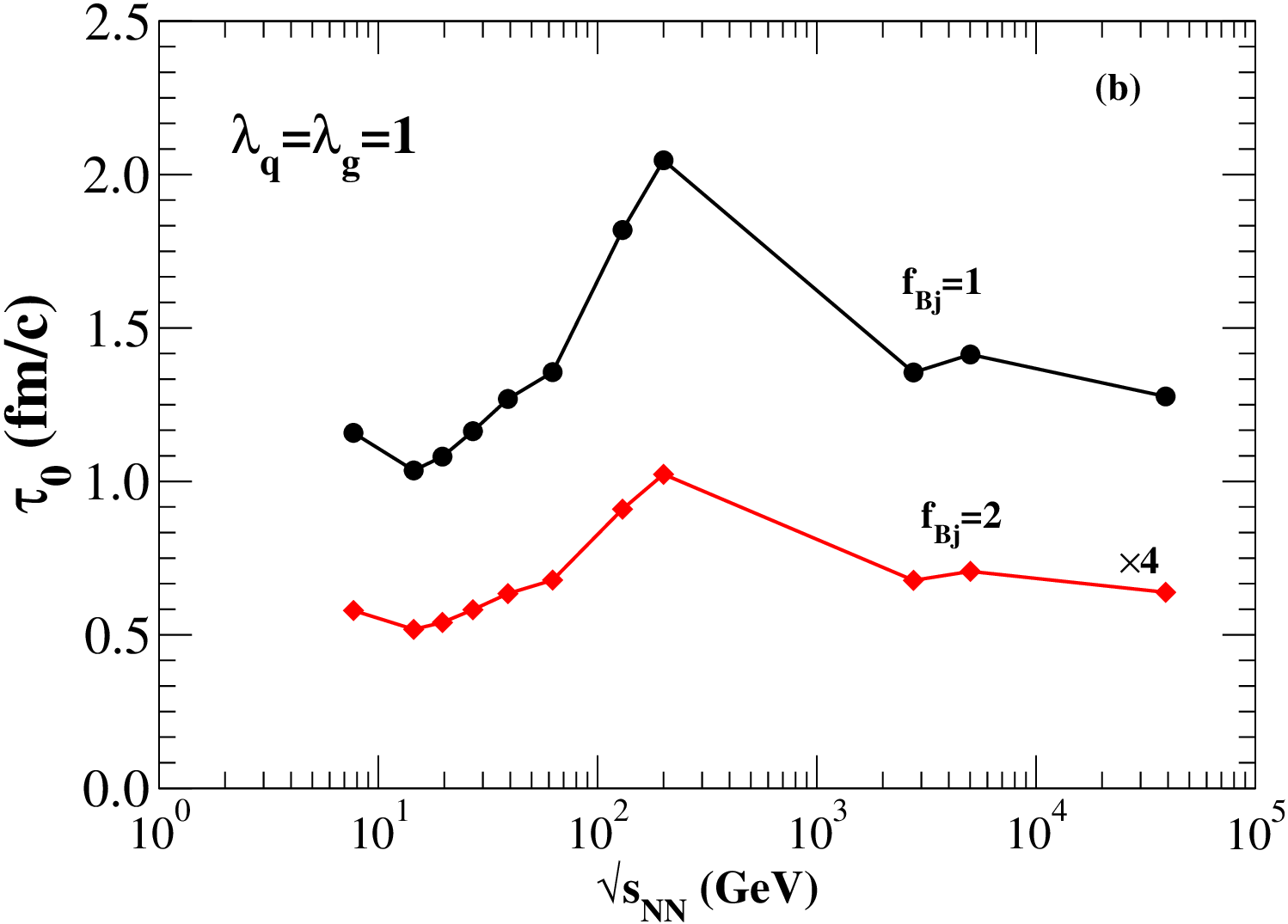}}
\caption{(Colour online) The initial temperature and formation time for energies at RHIC, LHC and FCC under the assumption of the formation of a thermally and chemically equilibrated plasma. The correction factor for the Bj\"{o}rken energy density is assumed to be equal to 1 or 2.}
\label{t_tau}
\end{figure}
As a first step we assume that a thermally and chemically equilibrated quark gluon plasma is formed in these collisions at a temperature $T_0$ at proper time $\tau_0$. This immediately implies that
\begin{equation}
\lambda_q(\tau_0)=\lambda_g(\tau_0)=1.
\end{equation} 
The rate equations~(Eq.~\ref{master_long}) then ensure that these will remain constant and equal to one during
the expansion and cooling. 

The energy density and the number density for thermally and chemically equilibrated plasma, with negligible net-baryons can be written as, 
\begin{equation}
\epsilon(\tau_0)=\left [ a_2 + 2 b_2 \right ] T_0^4
\end{equation}
and
\begin{equation}
n(\tau_0)= \left [  a_1 + 2 b_1 \right ] T_0^3
\end{equation}
so that
\begin{eqnarray}
T_0 &=& f_{\text{Bj}}\left [ \frac{a_1+2~b_1}{a_2+2~b_2} \right ]
\frac{dE_T/dy}{dN/dy}\nonumber\\
& \approx & 0.337 f_{\text{Bj}} \frac{dE_T/dy}{dN/dy}
\label{temp_1}
\end{eqnarray}
and
\begin{equation}
\tau_0=\frac{1}{\pi R_T^2  T_0^3(a_1+2 b_1)}\frac{dN}{dy}
\label{tau0_1}
\end{equation}
uniquely.

Actually this should not come as a suprise to us. The Eq.~\ref{temp_1} simply implies that the energy per
particle for a system of thermallized mass-less partons is $\approx 2.96 T_0$ (see also Ref.~\cite{kari}).
The crucial step that we take in this work is to use this knowledge along with the experimentally
measured rapidity density of produced particles to explore the formation time $\tau_0$  (Eq.~\ref{tau0_1})
and extent of chemical equilibration (see also later).

Before proceeding we recall that the Eq.~\ref{tau0_1} has often been used~\cite{hwa}
to estimate the initial temperature with {\em an assumed value for $\tau_0$}.

We give the results of this estimate for the data obtained at RHIC energies during a beam energy scan~\cite{rhic} and LHC~\cite{lhc} energy (2.76 ATeV) for the most central collision
(see Fig.~\ref{t_tau}).  The results for LHC (5.02 ATeV) and Future Circular Collider (39 ATeV) are also indicated based on estimated multiplicity and transverse energy rapidity density~\cite{fcc}. We show our results for two choices of the correction factor $f_{\text{Bj}}$, namely, 1 and 2, applied
at all the energies under consideration.

If we neglect the correction factor $f_{\textrm{Bj}}$, we find that the formation time of  the
plasma for the energies under consideration is of the order of 1--2 fm/$c$. The initial
temperatures at all energies are found to be in excess of 160 MeV, the putative value
for the (critical temperature for) quark-hadron transition. 

A rich structure is seen to emerge in the variation of the initial time ($\tau_0$) which rises from a value of $\approx$ 1 fm/$c$ at lowest energies, to a little more than 2 fm/$c$ at the top RHIC energy
and a similar drop to about 1.3--1.4 fm/$c$ at higher energies. This trend of rise and fall is also  seen when the large correction factor of 2 is used. 

At what energy  does it peak and where does the turn-over start?

We also note that the initial temperature in both the cases rises monotonically but logarithmically with the centre of mass energy.  One possible explanation for this behaviour could be an  increasing role played by the saturation of the partons at higher collision energies and an increase in their momenta.

Of-course this trend can be off-set or it can even be completely different, depending on the variation of 
 $f_{\textrm{Bj}}$ 
with the energy of the collision (see  also later).
\begin{figure}
\centerline{\includegraphics*[width=8.0 cm]{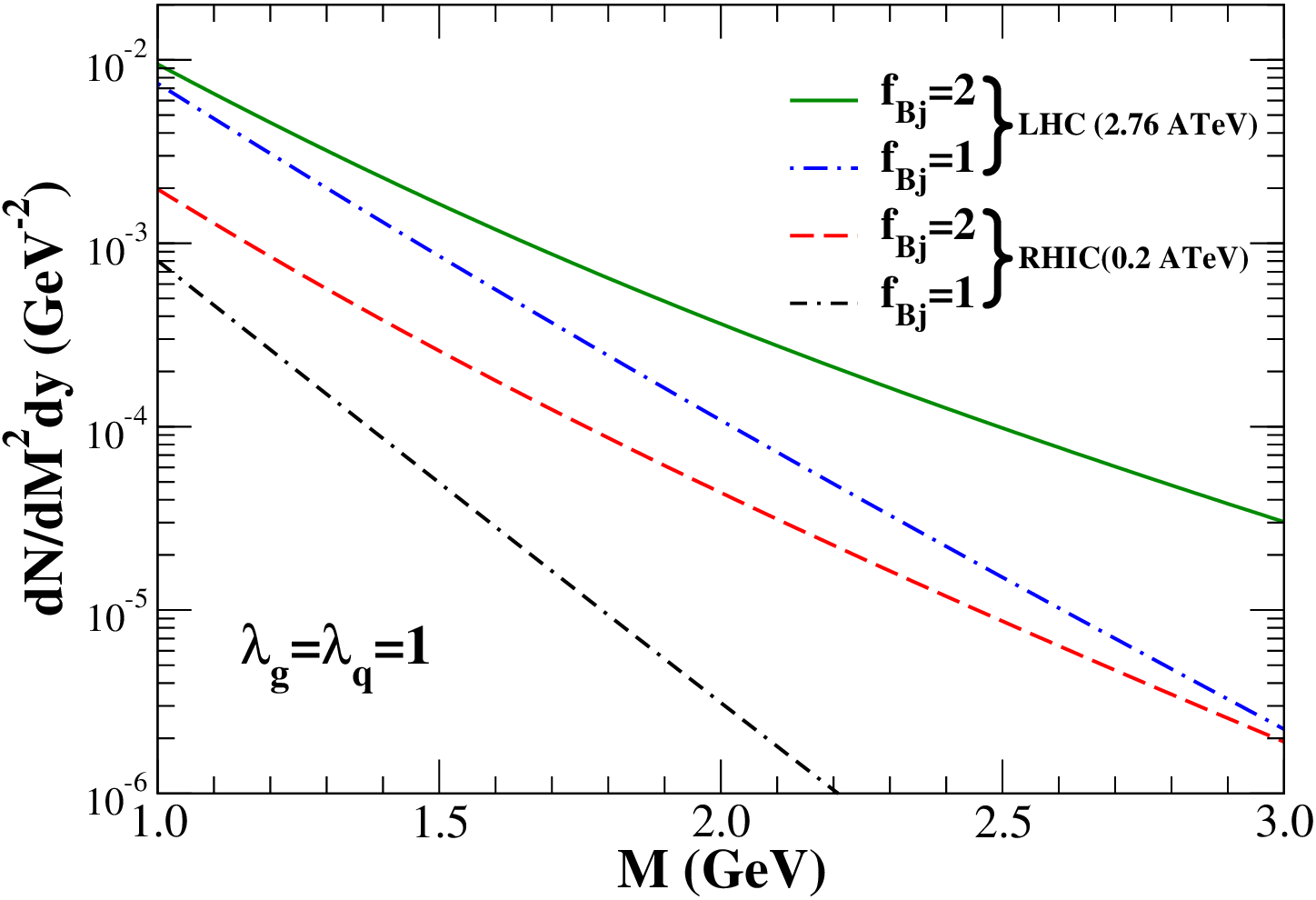}}
\caption{(Colour online)  Production of intermediate mass dileptons 
from quark-antiquark annihilation at RHIC (200 AGeV) and LHC (2.76 ATeV) energies
for $f_{\textrm{Bj}}$ equal to 1 and 2 respectively and 
$\lambda_q(\tau_0)=\lambda_g(\tau_0)= 1$. }
\label{dil_lamb1}
\end{figure}
\begin{figure}
\centerline{\includegraphics*[width=8.0 cm]{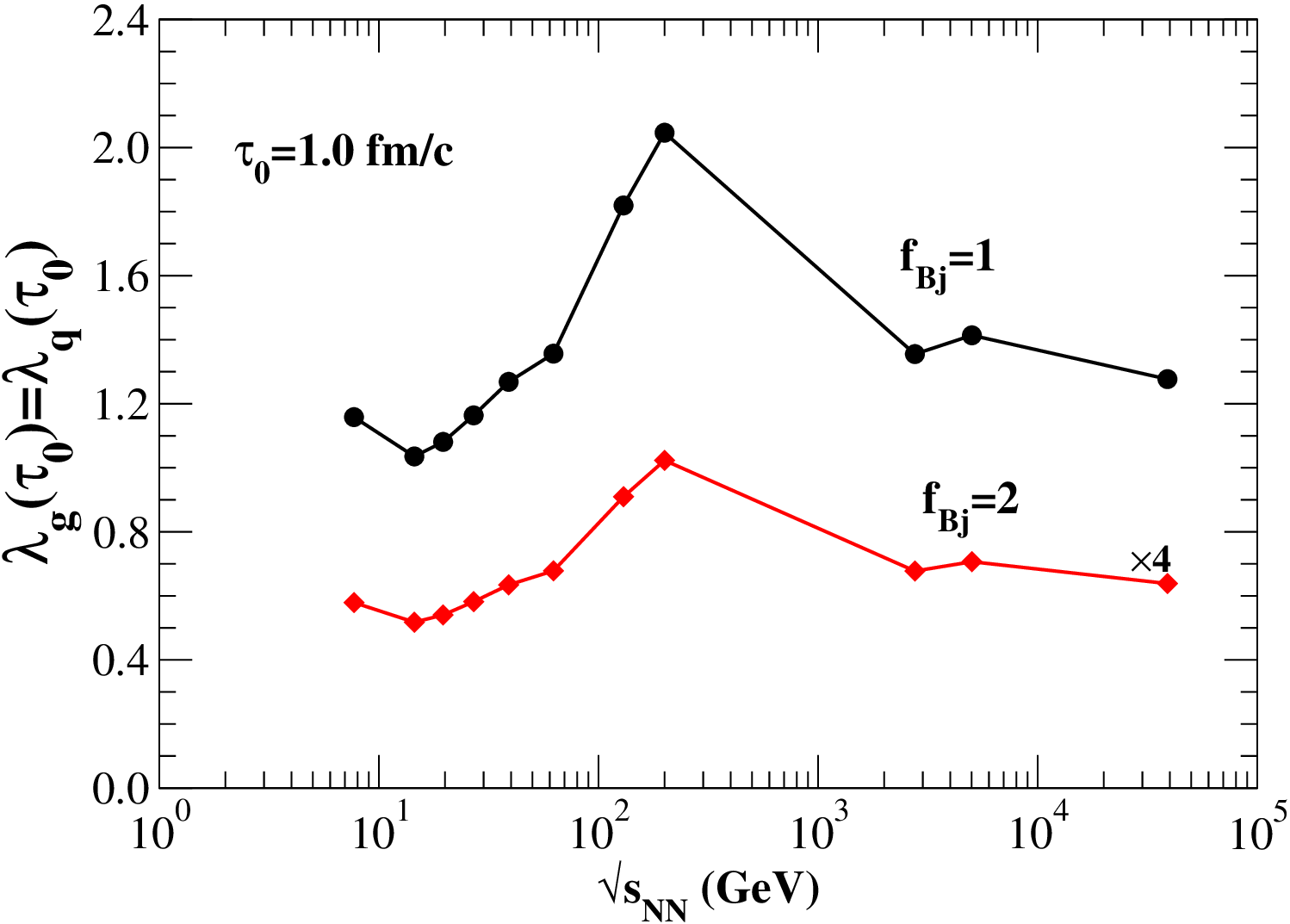}}
\centerline{\includegraphics*[width=8.0 cm]{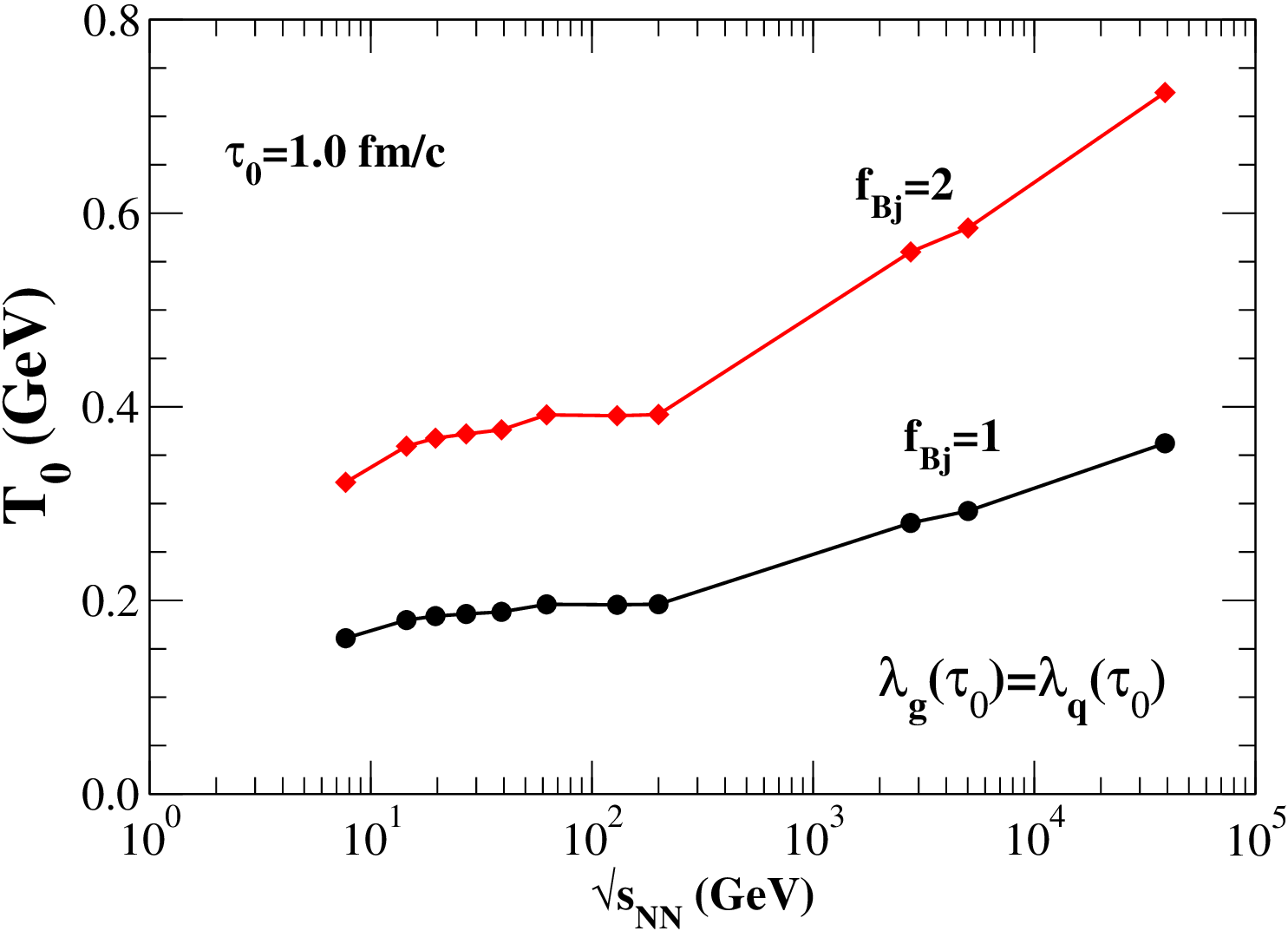}}
\caption{(Colour online) The initial fugacity and temperature 
for energies at RHIC, LHC and FCC under the 
assumption of the formation of a thermally equilibrated plasma at $\tau_0=$ 1 fm/$c$.
The correction factor for the Bj\"{o}rken energy density is assumed to be equal to
1 or 2.}
\label{chem-equal-1fm}
\end{figure}

The importance of including/neglecting the correction factor $f_\text{Bj}$ is immediately and abundantly clear. Thus we see that assuming a value of 2 for this correction factor leads to an increase in the initial temperature ($T_0$) by a factor of 2 and  a decrease in the formation time ($\tau_0$) by a factor of eight.

In Fig.~\ref{dil_lamb1} we give the invariant mass distribution of intermediate mass
dileptons from quark-antiquark annihilation, following the treatment of Ref.~\cite{kkmm}, for RHIC energy (200 AGeV) and at LHC energy (2.76 ATeV) from the QGP phase. As expected, we see that the results are quite sensitive to the estimates for the corrections to Bj\"{o}rken energy density.
Similar differences were found for the FCC energy.

\begin{figure}
\centerline{\includegraphics*[width=8.0 cm]{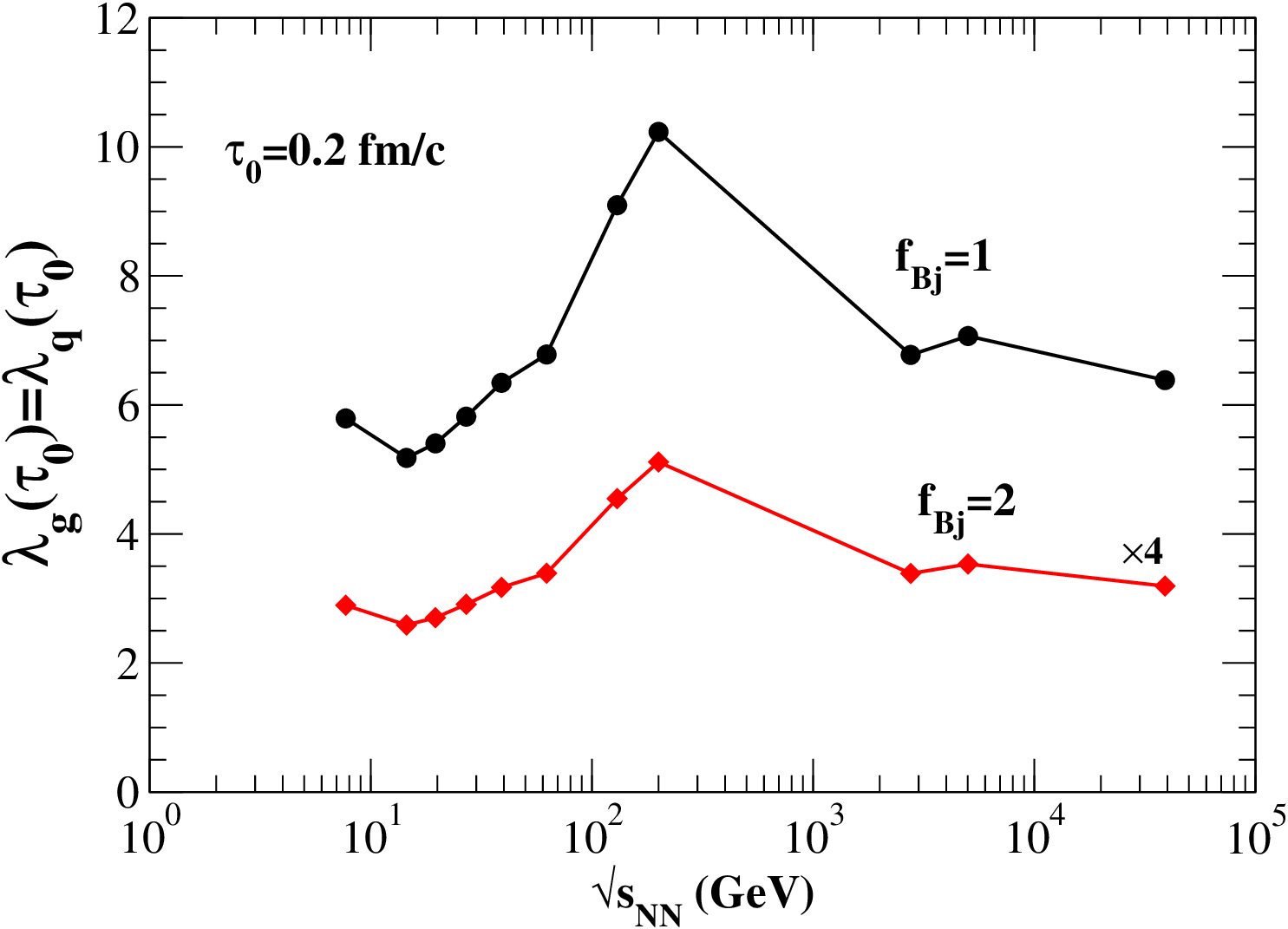}}
\centerline{\includegraphics*[width=8.0 cm]{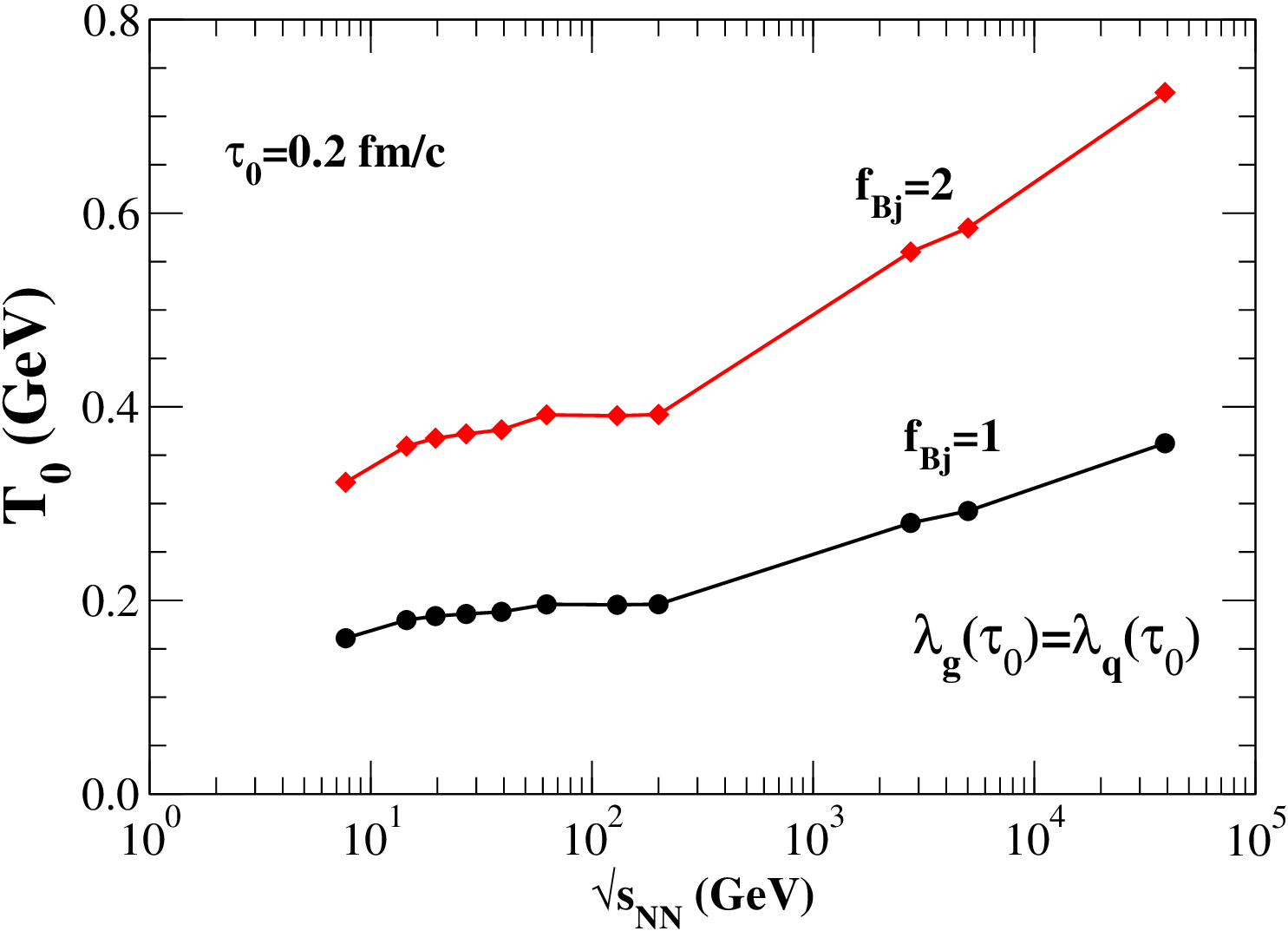}}
\caption{(Colour online) The initial fugacity and temperature 
for energies at RHIC, LHC and FCC under the 
assumption of the formation of a thermally equilibrated plasma at $\tau_0=$ 0.2 fm/$c$.
The correction factor for the Bj\"{o}rken energy density is assumed to be equal to
1 or 2. Note that the results for the initial temperature are identical to
those in Fig.\ref{chem-equal-1fm} and are repeated here to emphasize this point.}
\label{chem-equal-0.2fm}
\end{figure}

\subsection{Thermally equilibrated and chemically equilibrating quark gluon plasma}
We note that with the neglect of the net baryons in the central rapidity region, we need four parameters to describe a thermally equilibrated quark gluon plasma, $\lambda_g$, $\lambda_q$, $T_0$, and $\tau_0$. 
In the above we have seen that the assumption of chemical equilibration provides unique values for $T_0$ and 
$\tau_0$ once the correction factor $f_{\text{Bj}}$ is specified. 

\subsubsection{$\lambda_g(\tau_0) = \lambda_q(\tau_0)$}
In the following we relax this condition. As a first step we assume that the quarks and gluons are equilibrated to the same extent at the beginning, i.e.,
\begin{equation}
\lambda_q(\tau_0)=\lambda_g(\tau_0).
\end{equation} 

This fixes the initial temperature uniquely according to 
Eq.~\ref{temp_1}.

 However as mentioned earlier, now the formation time  has to be determined from
\begin{equation}
\lambda_g(\tau_0) \tau_0=\frac{1}{\pi R_T^2  T_0^3(a_1+2 b_1)} \frac{dN}{dy}.
\label{tau0_2}
\end{equation}

\begin{figure}
\centerline{\includegraphics*[width=8.0 cm]{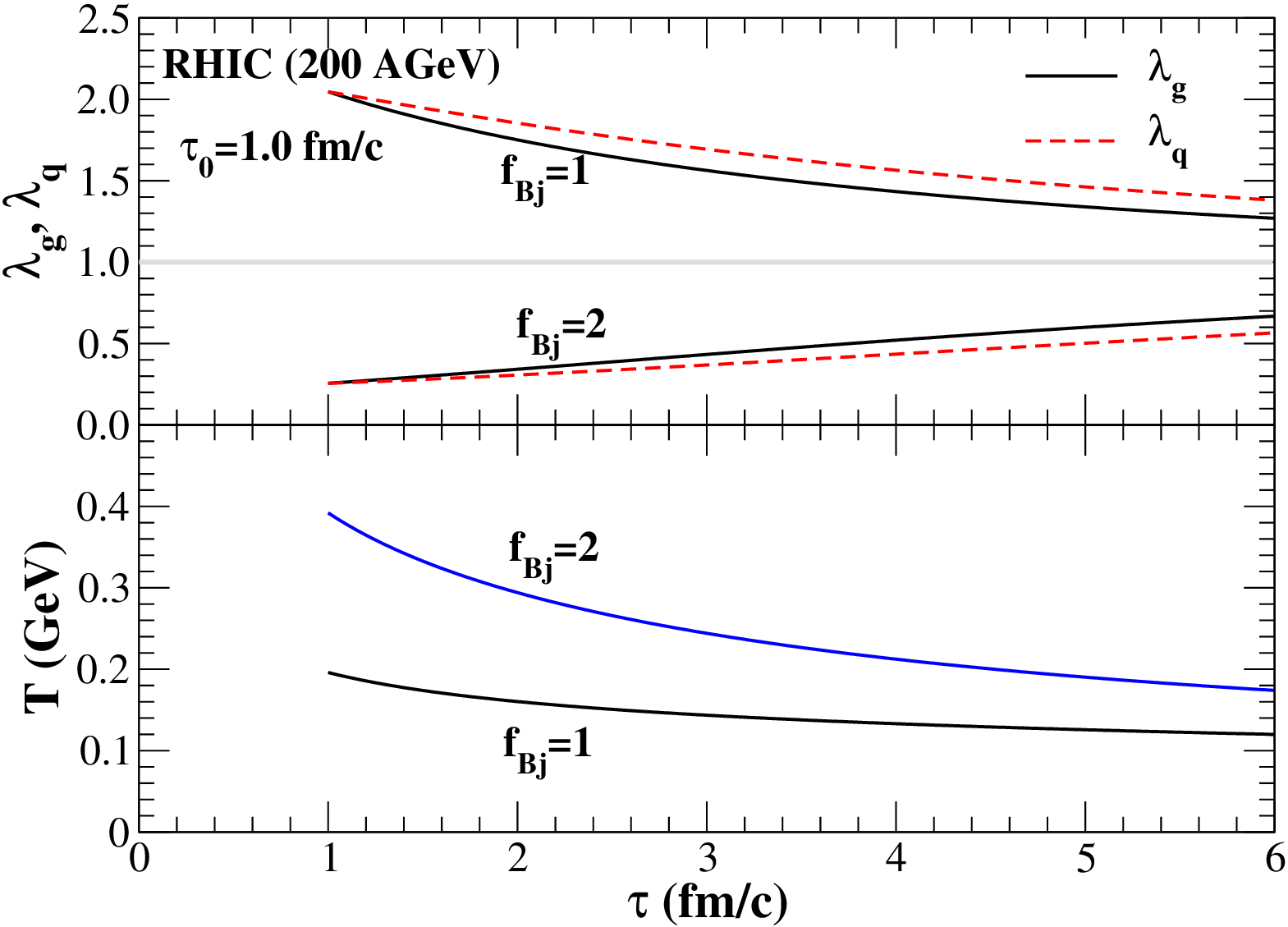}}
\centerline{\includegraphics*[width=8.0 cm]{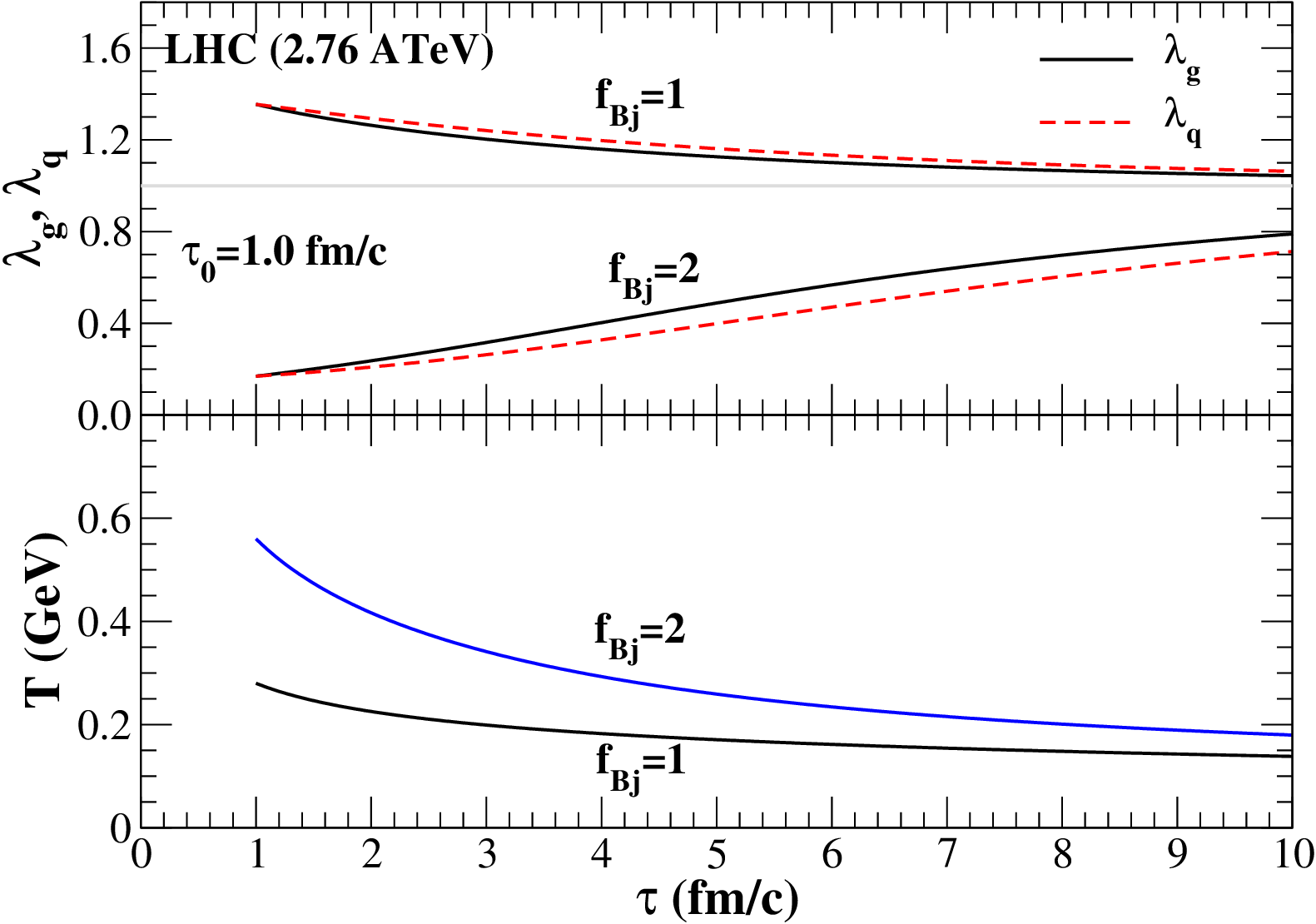}}
\centerline{\includegraphics*[width=8.0 cm]{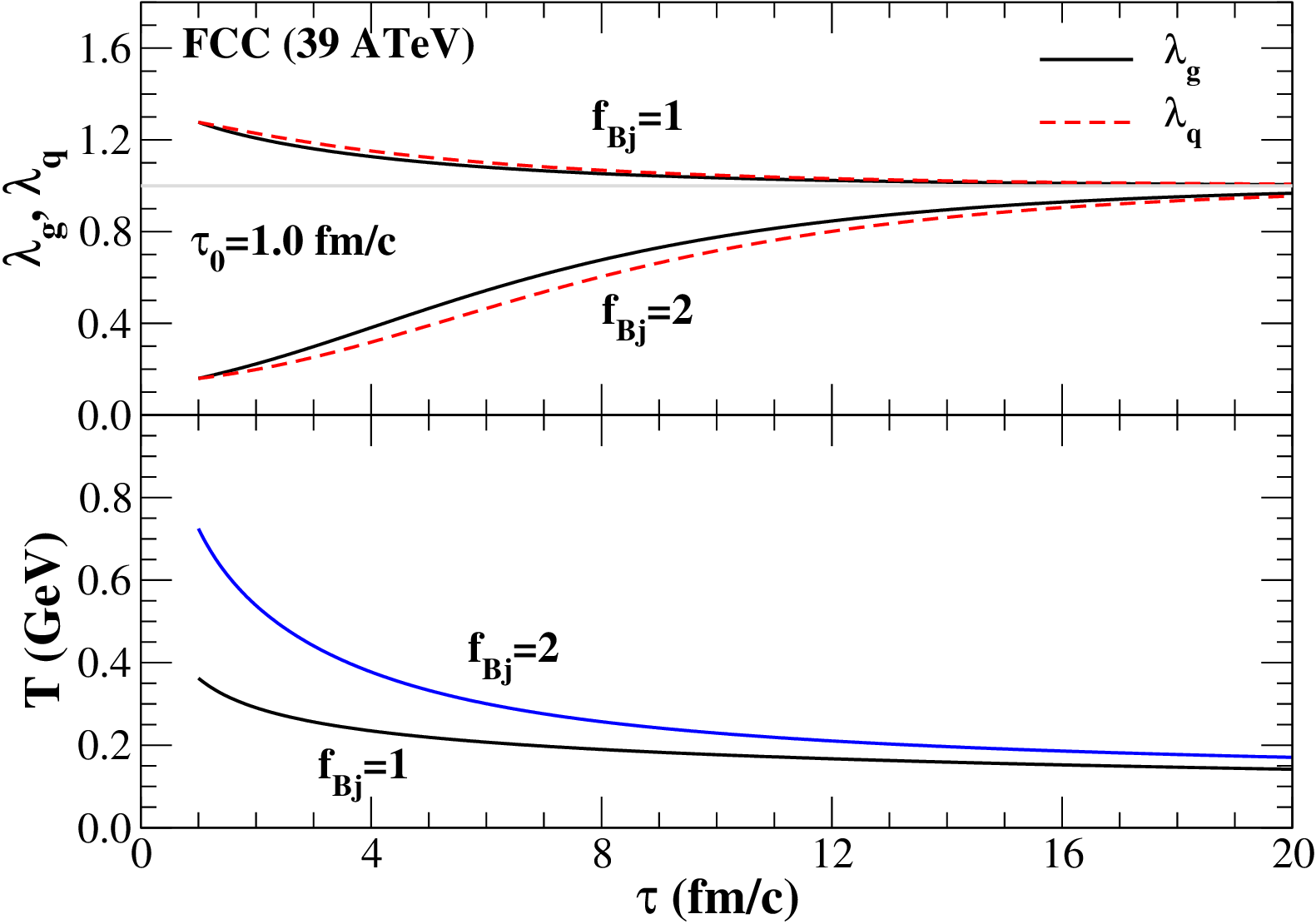}}
\caption{(Colour online) The time evolution of fugacities and temperature
at RHIC (200 AGeV), LHC (2.76 ATeV) and FCC (39 ATeV) with the
assumption of formation time, $\tau_0=$ 1.0 fm/$c$ and 
the correction factor,
$f_{\textrm {Bj}}=$1 and 2.}
\label{fbj_lam_tau1}
\end{figure}

\begin{figure}
\centerline{\includegraphics*[width=8.0 cm]{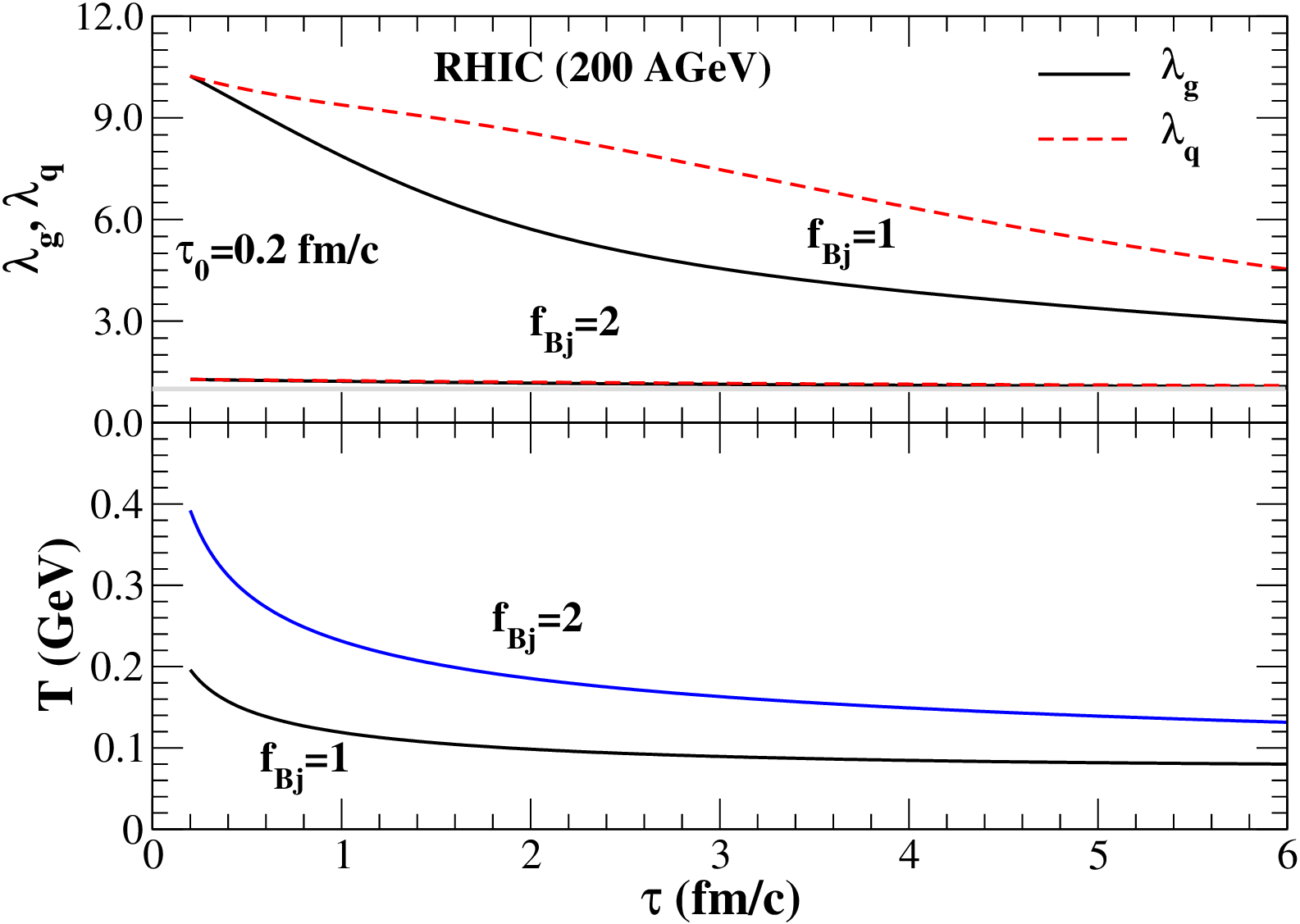}}
\centerline{\includegraphics*[width=8.0 cm]{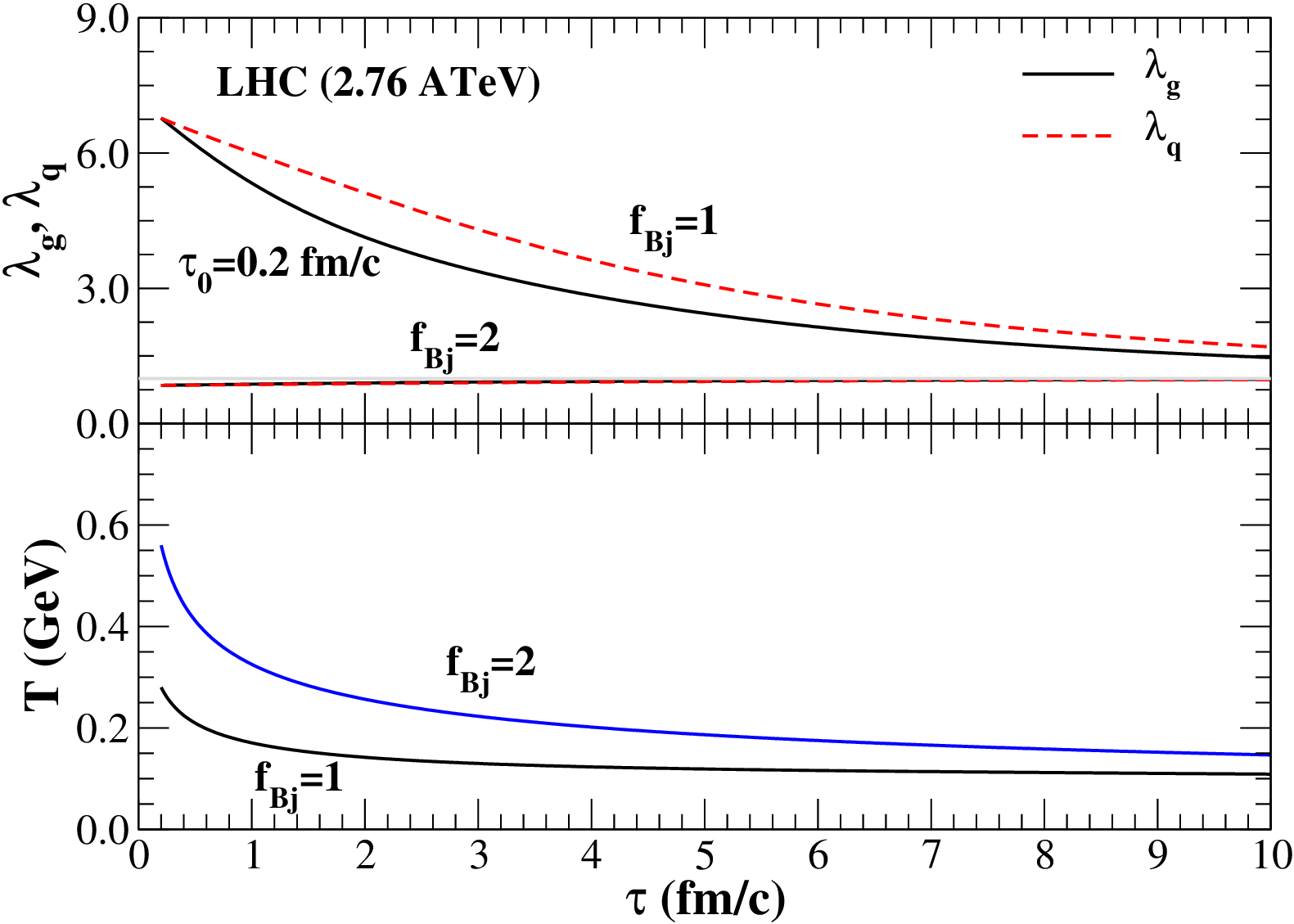}}
\centerline{\includegraphics*[width=8.0 cm]{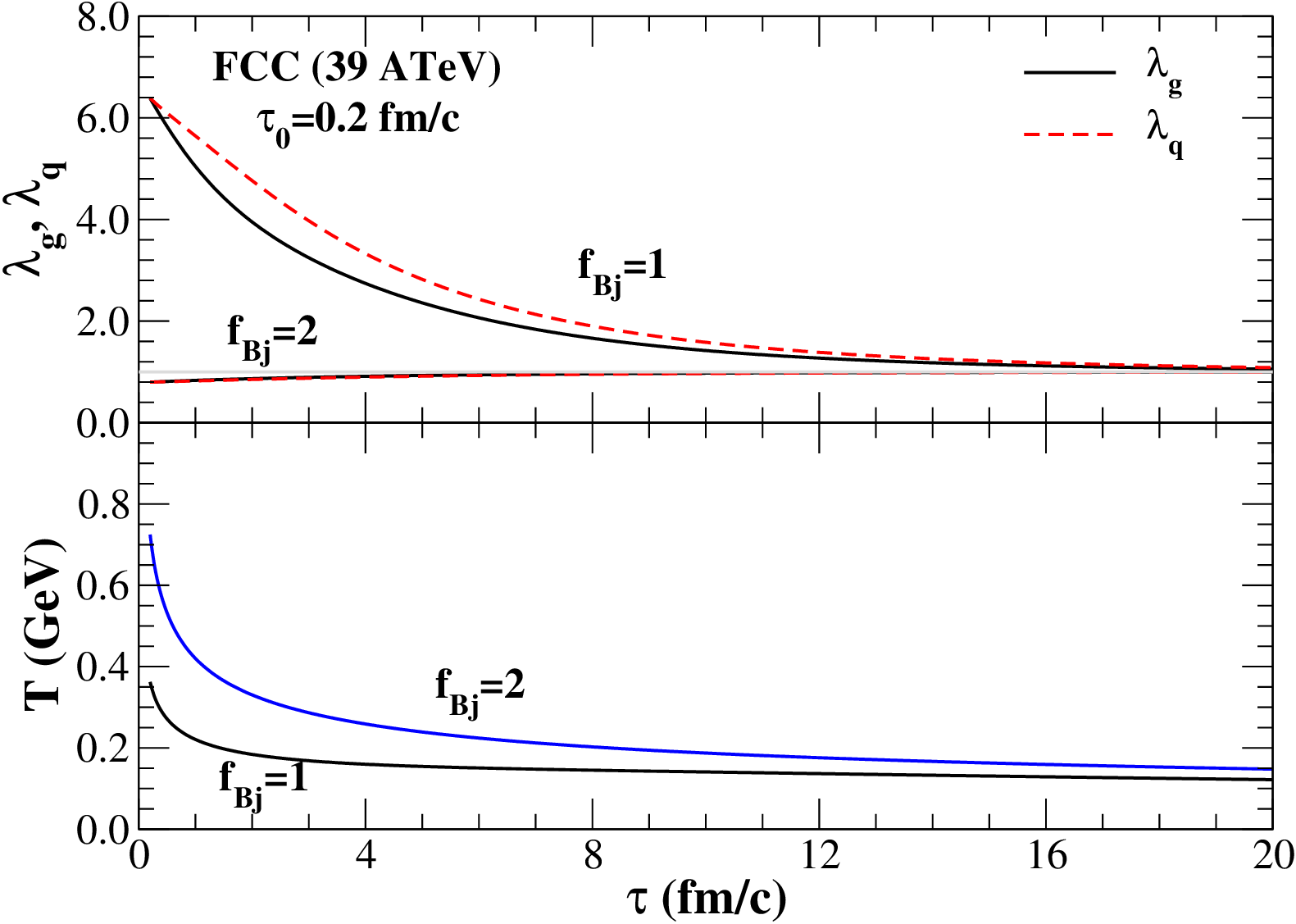}}
\caption{(Colour online) The time evolution of fugacities and temperature
at RHIC (200 AGeV), LHC (2.76 ATeV) and FCC (39 ATeV) with the
assumption of formation time, $\tau_0=$ 0.2 fm/$c$ and 
the correction factor,
$f_{\textrm {Bj}}=$1 and 2.}
\label{fbj_lam_tau0.2}
\end{figure}
This provides that if $\tau_0$ is large, the extent of chemical equilibration will be
small. However once again, the correction factor $f_{\textrm{Bj}}$  plays a significant role, as the product $\lambda_g(\tau_0) \tau_0$ is proportional to $1/f_{\textrm {Bj}}^3$. 

While one expects smaller $\tau_0$ for higher incident energies from several considerations, e.g., increased $\langle p_T \rangle$ for higher incident energies implying higher initial temperatures\cite{kms} with  
 $\tau_0 \propto 1/3T_0$ or higher saturation momenta $p_{\textrm{sat}}$ for higher incident energies~\cite{kari}, so that $\tau_0 \propto 1/p_{\textrm{sat}}$, its precise value remains indefinite.

In order to encompass these limits, we discuss our results for two typical values for $\tau_0$ namely, 1 fm/$c$ (Fig.~\ref{chem-equal-1fm}) and 0.2 fm/$c$ (Fig.~\ref{chem-equal-0.2fm}), favoured  by various groups for initiating the hydrodynamic calculations. Results
for any arbitrary value of $\tau_0$ can be easily calculated. 

Now the crucial dependence of the extent of chemical equilibration on the  correction to Bj\"{o}rken energy density becomes even more interesting. We see (Fig.~\ref{chem-equal-1fm}) that for $f_{\textrm{Bj}}=1$ and $\tau_0=$ 1 fm/$c$, the quarks and gluons are in a state of super-saturation for all energies under consideration.

However if $f_{\textrm{Bj}}$ is chosen as equal to 2, while the energy density increases by a factor of 2,
the plasma at  all the energies  is under-saturated as the initial temperature increases by a factor of 2,
and the initial fugacities $\lambda_g(\tau_0)=\lambda_q(\tau_0)$ decrease by a factor of 8 (see Fig.~\ref{chem-equal-1fm}).

Looking at Fig.~\ref{chem-equal-0.2fm}, we find that we have more surprises in store for us. For $f_{\textrm{Bj}}=1$ and $\tau_0=$ 0.2 fm/$c$ we find that plasma is grossly oversaturated
at all energies, while it is moderately undersaturated for $f_{\textrm{Bj}}=2$.
 
We now look at the time evolution of the fugacities. For the sake of brevity, we consider only three cases,
RHIC top energy, LHC (2.76 ATeV) and FCC (39 ATeV) for the two cases discussed above (Fig.~\ref{fbj_lam_tau1}).  

We note that the super-saturation of quarks and gluons noted at all the energies decreases with passage of time and approaches chemical equilibration. However, even towards the end of QGP phase, the plasma at RHIC (200 AGeV)
remains away from chemical equilibrium, remaining over-saturated for $f_{\textrm {Bj}}=$ 1 and remaining under-saturated for $f_{\textrm {Bj}}=$ 2.

The results for the LHC(2.76 ATeV) are also quite interesting in the sense that the plasma approaches chemical equilibrium for $f_{\textrm {Bj}}=$ 1 but remains moderately under-saturated when it is taken as 2. 

The plasma at FCC approaches chemical equilibration for both cases. 


Next we assume that the formation time $\tau_0$ = 0.2 fm/$c$. Calculations similar to that discussed above are given in Fig.~\ref{fbj_lam_tau0.2}.

We have already noted that the initial temperatures are not altered  by reduction of the formation time in our description. When $f_{\textrm {Bj}}$ is taken as 1 and the formation time $\tau_0$ as
0.2 fm/$c$, then the plasma at all energies is necessarily produced in a highly super-saturated form, though it attains the chemical equilibration towards the end of the QGP phase for the higher energies. We must add that it is not clear whether such a highly super-saturated plasma would be stable.

If however $f_{\textrm {Bj}}$ is taken as 2, the plasma at all the energies under consideration is produced very close to a chemical equilibration. If confirmed, this lends a very strong  support to all the calculations of electromagnetic radiation where a chemically equilibrated quark gluon plasma is 
assumed to be formed in relativistic heavy ion collisions, with a formation time of the order of 0.2 fm/$c$.

\begin{figure}
\centerline{\includegraphics*[width=8.0 cm]{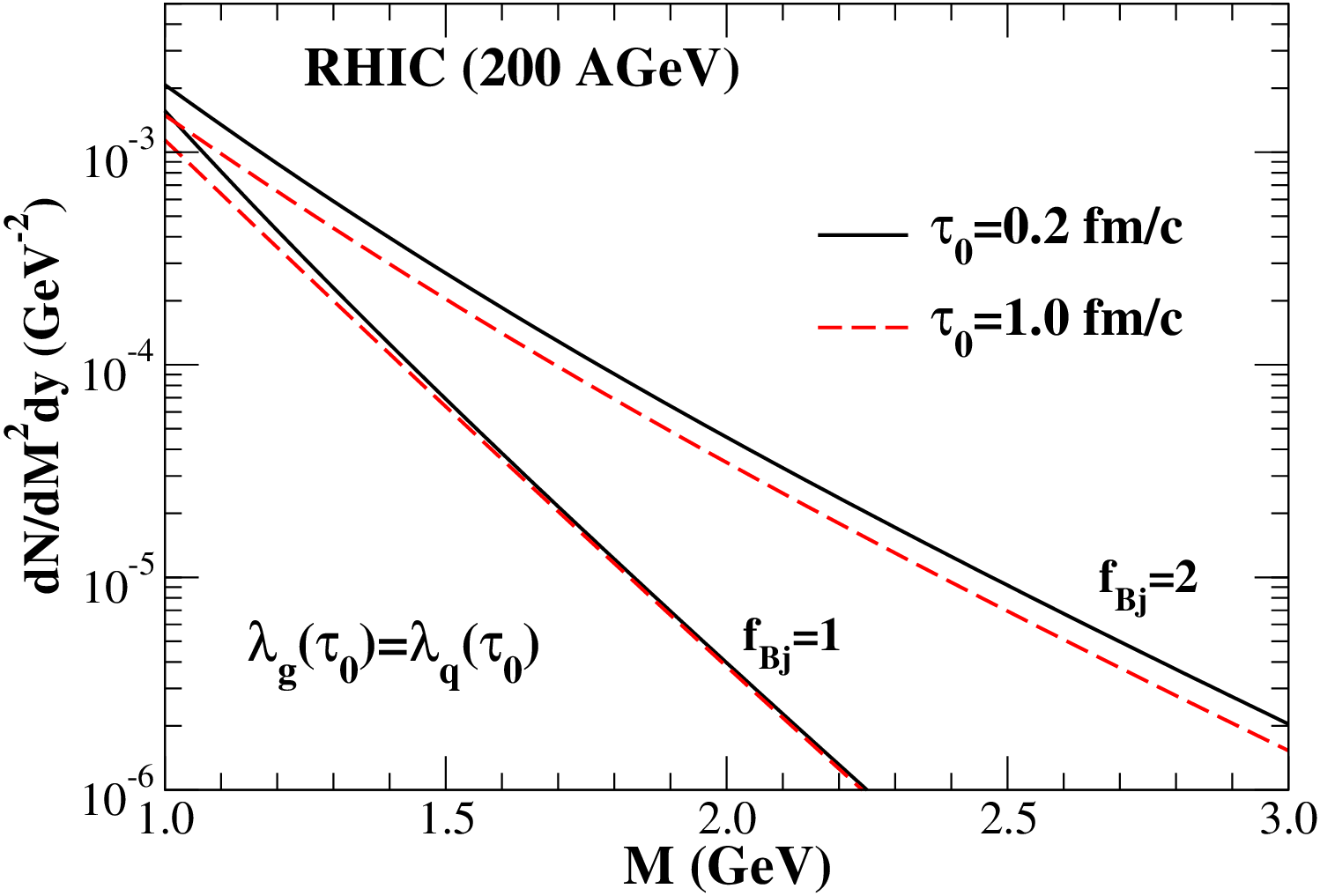}}
\centerline{\includegraphics*[width=8.0 cm]{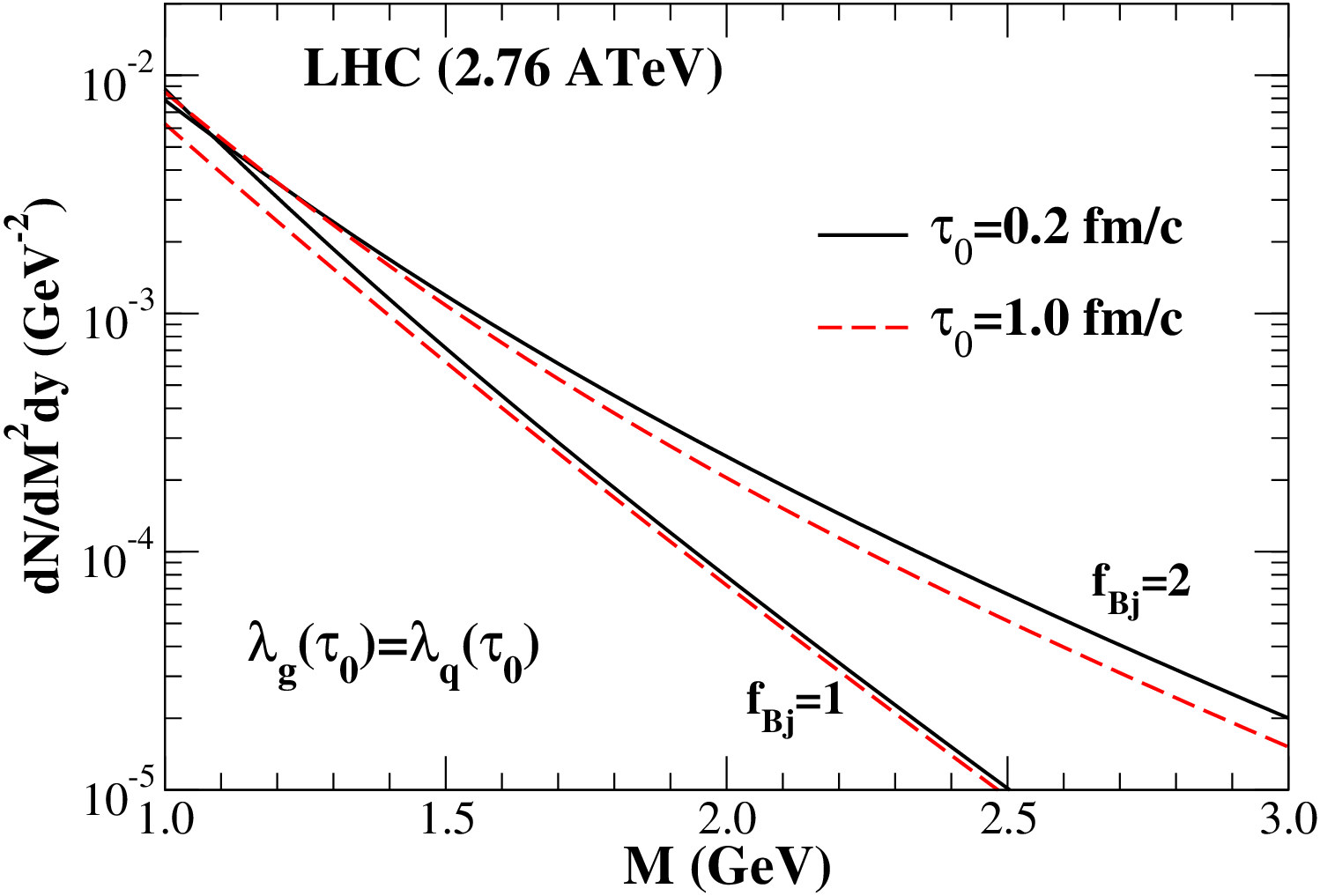}}
\centerline{\includegraphics*[width=8.0 cm]{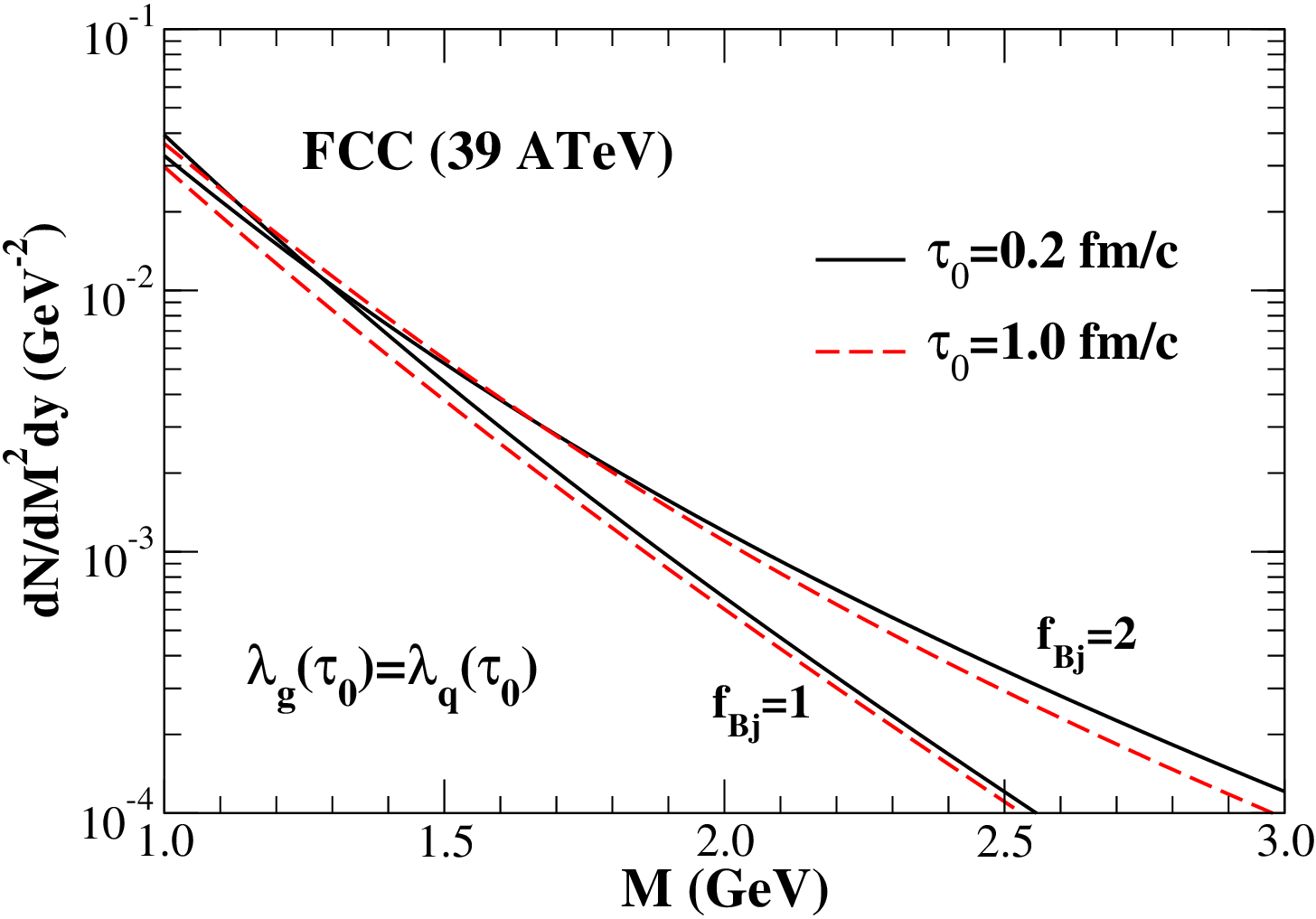}}
\caption{(Colour online) The invariant mass 
distribution of intermediate mass dileptons radiated from
nucleus-nucleus collision at RHIC (200 AGeV), LHC (2.76 ATeV) 
and FCC (39 ATeV) when of formation times
is reduced from 1 fm/$c$ to 0.2 fm/$c$ and the correction
factor $f_{\textrm {Bj}}$ increased from 1 to 2.}
\label{dil_lam}
\end{figure}

Can the radiation of dileptons having intermediate mass be used to  distinguish the different evolution scenarios discussed here? In order to assess this we show our results for the above cases
in Fig.~\ref{dil_lam}.

We find that for all the cases under consideration, the production of dileptons having intermediate mass is only marginally affected by reduction of the formation time from 1.0 fm/$c$ to 0.2 fm/$c$ for
a given $f_{\textrm {Bj}}$, even though we have seen above that it drastically alters the phase-space
occupancy of the partons in the plasma through the entire history of evolution!  

However the results are found to be extremely sensitive to the value of $f_{\textrm {Bj}}$.
It is already known (see Ref.~\cite{munshi}) that this result would not be greatly altered
by inclusion of transverse expansion of the plasma.

This  could be quite useful in determining the precise value of $f_{\textrm {Bj}}$ and the
initial energy density.

\subsubsection{$\lambda_g(\tau_0) \neq \lambda_q(\tau_0)$}

It has been discussed for a long time that the initial state of the plasma could be rich in 
gluons~\cite{hot_glue}, because $\sigma_{gg} > \sigma_{qg} > \sigma_{qq}$ and number of gluons in the nucleons could be much larger than the number of quarks. Gluon multiplication via $gg \rightarrow ggg$ and quark fragmentations $q \rightarrow qg$
also lends support to this supposition. Several studies reported earlier~\cite{biro,munshi,sspc,duncan,kari}
have utilized this condition. The colour gluon condensate initial state would also indicate this.

Thus, as an example, we assume that the initial fugacities are such that:

\begin{equation}
\lambda_q (\tau_0)= \alpha \lambda_g (\tau_0)
\end{equation}
and we take (somewhat arbitrarily), 
\begin{equation}
\alpha=0.2~.
\end{equation}

Proceeding as before we note that once again the initial temperature is uniquely fixed and the Eq.~\ref{temp_1}, reduces to
\begin{equation}
T_0 = f_{\text{Bj}}\left(\frac{a_1+2~\alpha b_1}{a_2+2~\alpha b_2}\right)
\frac{dE_T/dy}{dN/dy}~
\label{temp_2}
\end{equation}
Similarly, the initial time/fugacity will now be determined via:
\begin{equation}
\lambda_g(\tau_0) \tau_0=\frac{1}{\pi R_T^2  T_0^3(a_1+2 \alpha b_1)} \frac{dN}{dy}.
\label{tau0_3}
\end{equation}
\begin{figure}
\centerline{\includegraphics*[width=8.0 cm]{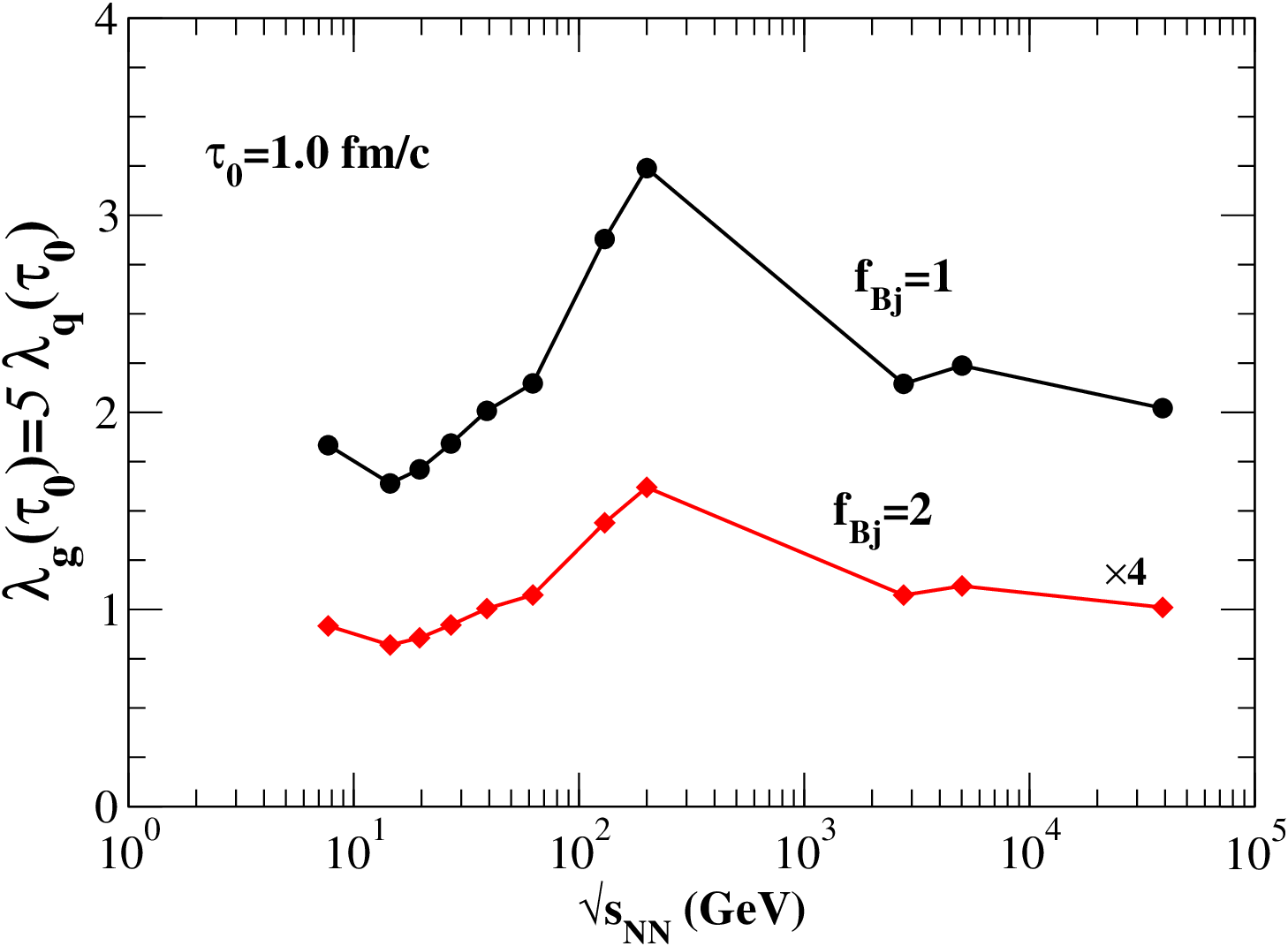}}
\centerline{\includegraphics*[width=8.0 cm]{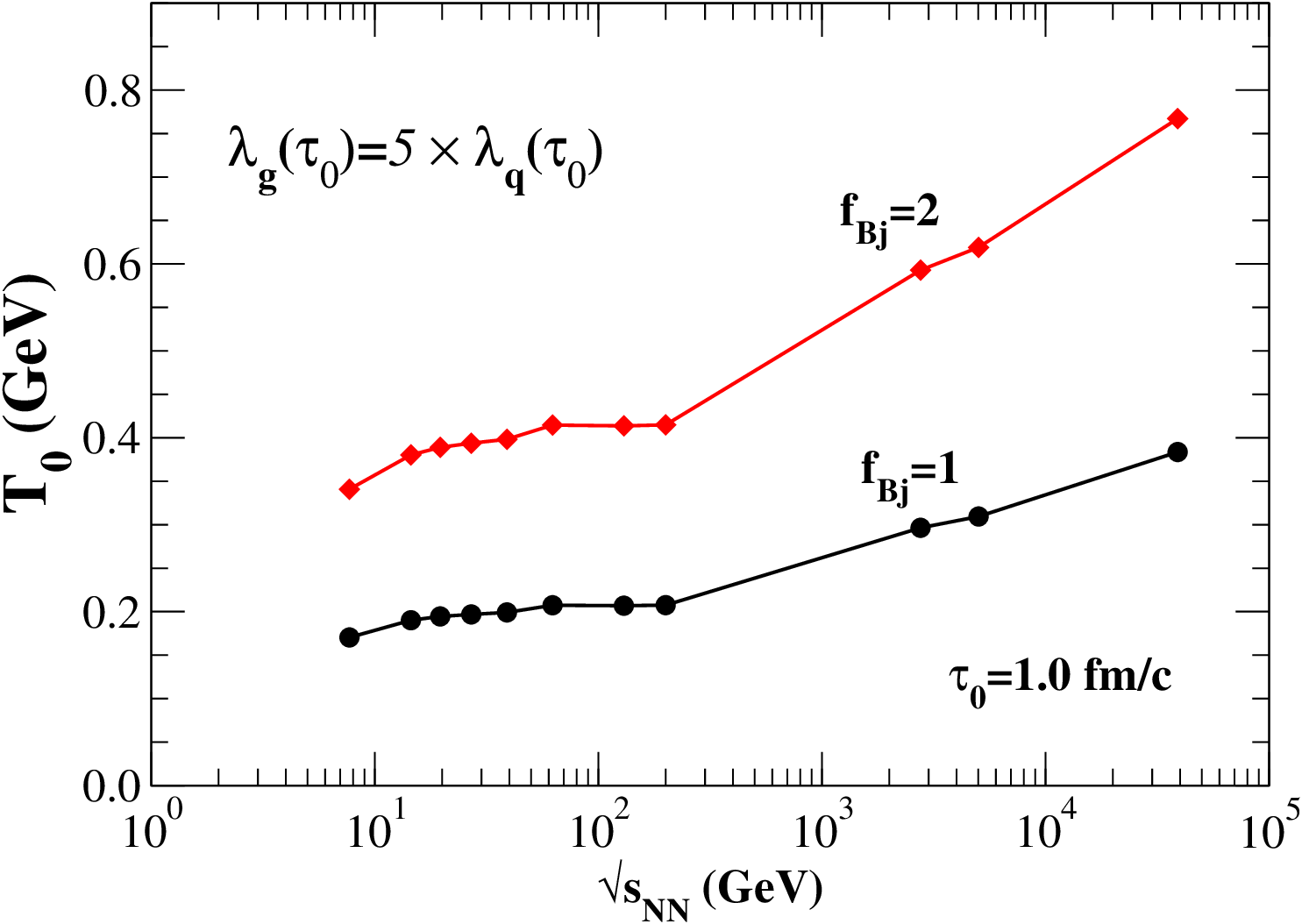}}
\caption{(Colour online) The initial fugacity and temperature 
 at RHIC, LHC and FCC energies under the 
assumption of the formation of a thermally equilibrated plasma at $\tau_0=$ 1.0 fm/$c$.
The correction factor for the Bj\"{o}rken energy density is assumed to be equal to
1 or 2. The extent of initial chemical equilibration is assumed to be such that 
$\lambda_g(\tau_0)= 5 \lambda_g(\tau_0)$.}
\label{chem-alpha-1fm}
\end{figure}

\begin{figure}
\centerline{\includegraphics*[width=8.0 cm]{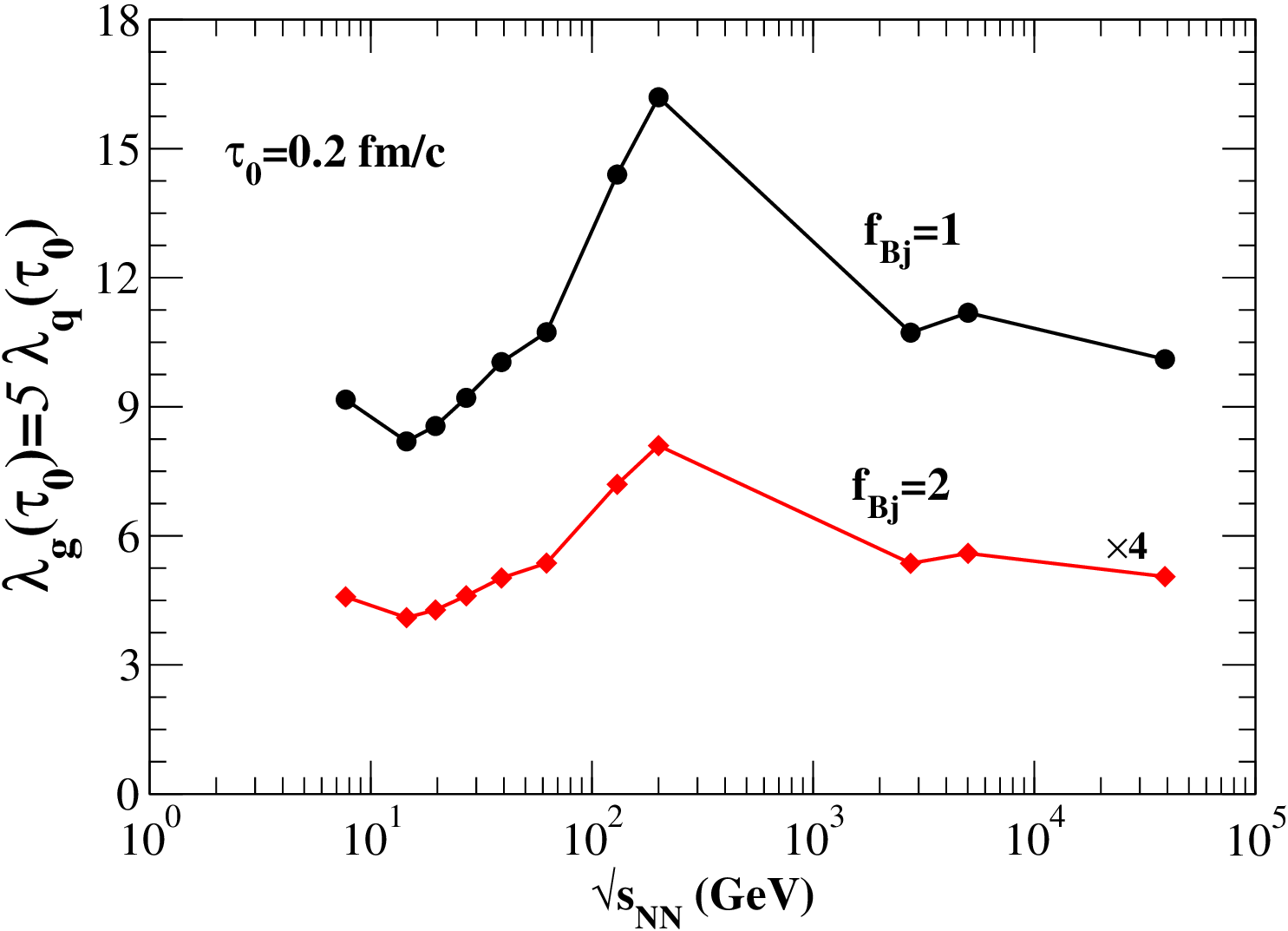}}
\centerline{\includegraphics*[width=8.0 cm]{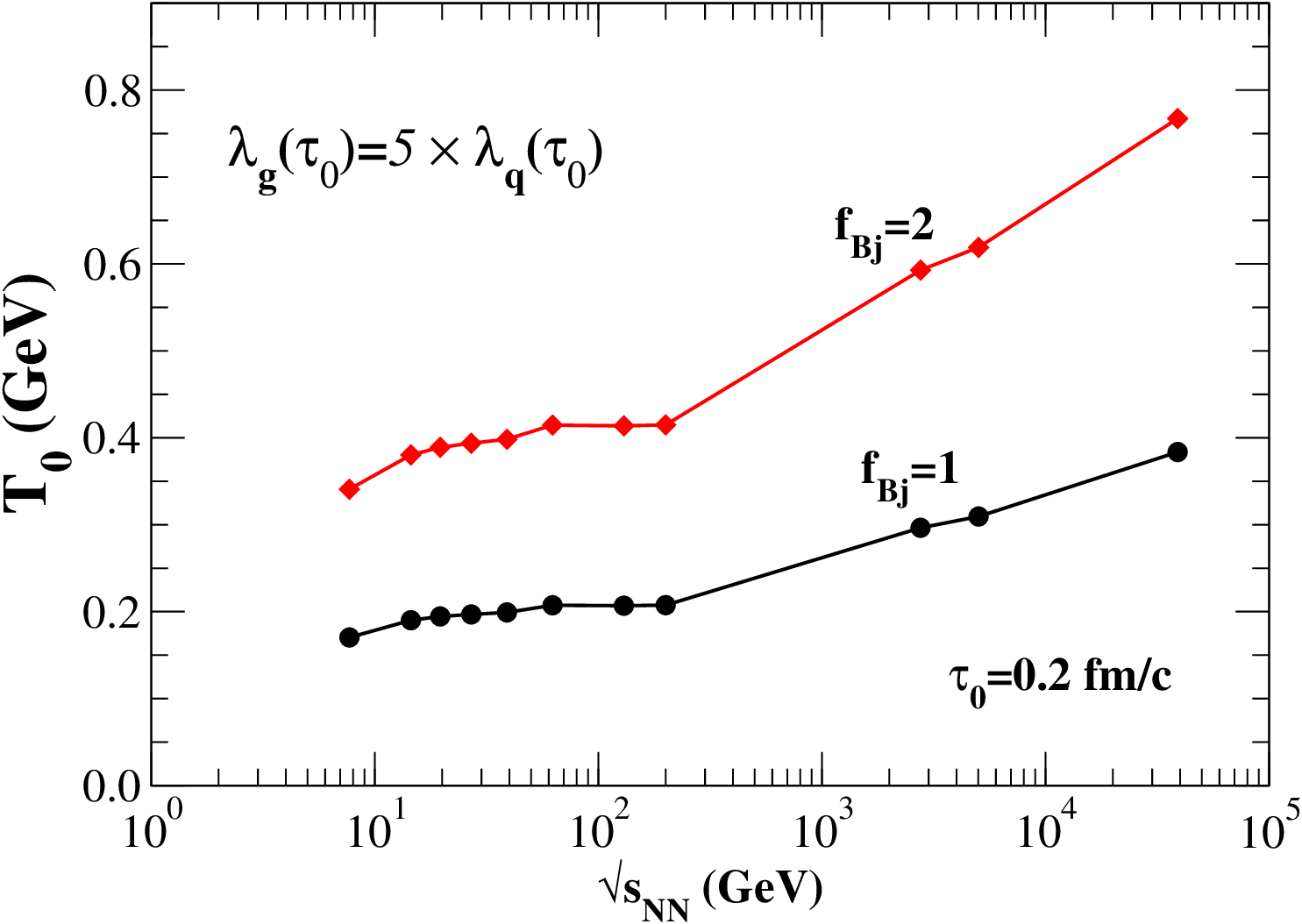}}
\caption{(Colour online) The initial fugacity and temperature at RHIC, LHC and FCC energies under the 
assumption of the formation of a thermally equilibrated plasma at $\tau_0=$ 0.2 fm/$c$.
The correction factor for the Bj\"{o}rken energy density is assumed to be equal to
1 or 2. The extent of initial chemical equilibration is assumed to be such that 
$\lambda_g(\tau_0)= 5 \lambda_g(\tau_0)$.}
\label{chem-alpha-0.2fm}
\end{figure}
We note that the fraction $(a_1 + 2 \alpha b_1)/(a_2 + 2 \alpha b_2)$ is $\approx$ 0.357 for $\alpha$ = 0.2 and thus the initial temperatures determined now will be about 6\% larger than that determined earlier 
using Eq.~\ref{temp_1}. In the extreme case of vanishing $\alpha$, this quantity is about 0.370 and will mean an increase of  about 10\% in the initial temperature, for the same energy density.

This as well as the smaller value for $\lambda_q$ are then compensated by an increase in the value for $\lambda_g$.

Thus we see (Fig.~\ref{chem-alpha-1fm}) that for $\tau_0=$ 1 fm/$c$ and $f_{\textrm {Bj}}=$ 1, as before
in Fig.~\ref{chem-equal-1fm} the initial temperature is more than 170 MeV at all the energies under consideration. The initial fugacity for gluons is also more than 1 at all energies and has a  peak at RHIC (200 AGeV). The quarks are under-saturated mostly because of the assumption that 
$\lambda_q(\tau_0)$ has a value which is one fifth of the value for $\lambda_g(\tau_0)$.

Once again we see that increasing the $f_{\textrm {Bj}}$ by a factor of two increases the initial temperature by a factor of two, (see Eq.~\ref{temp_2}). However, now the gluons as well as quarks are produced in a grossly under-saturated state of chemical equilibrium.

Assuming a formation time of 0.2 fm/$c$ and taking $f_{\textrm {Bj}}=$ 1 leads to the same initial temperature as before (and larger than 170 MeV) at all the energies under consideration and a grossly over-saturated  plasma-the over-saturation reaching a peak at the top RHIC energy, for gluons.  In fact even though we have assumed $\lambda_q=\lambda_g/5$ at the time of formation $\tau_0$, even the quarks are over-saturated with
$\lambda_q >$ 1 at all energies (Fig.~\ref{chem-alpha-0.2fm}). 

Enhancing the value for $f_{\textrm {Bj}}$ to 2, increases the initial temperature by a factor of 2 as before but now the plasma is moderately over-saturated for gluons and under-saturated for quarks at all energies.

\begin{figure}
\centerline{\includegraphics*[width=8.0 cm]{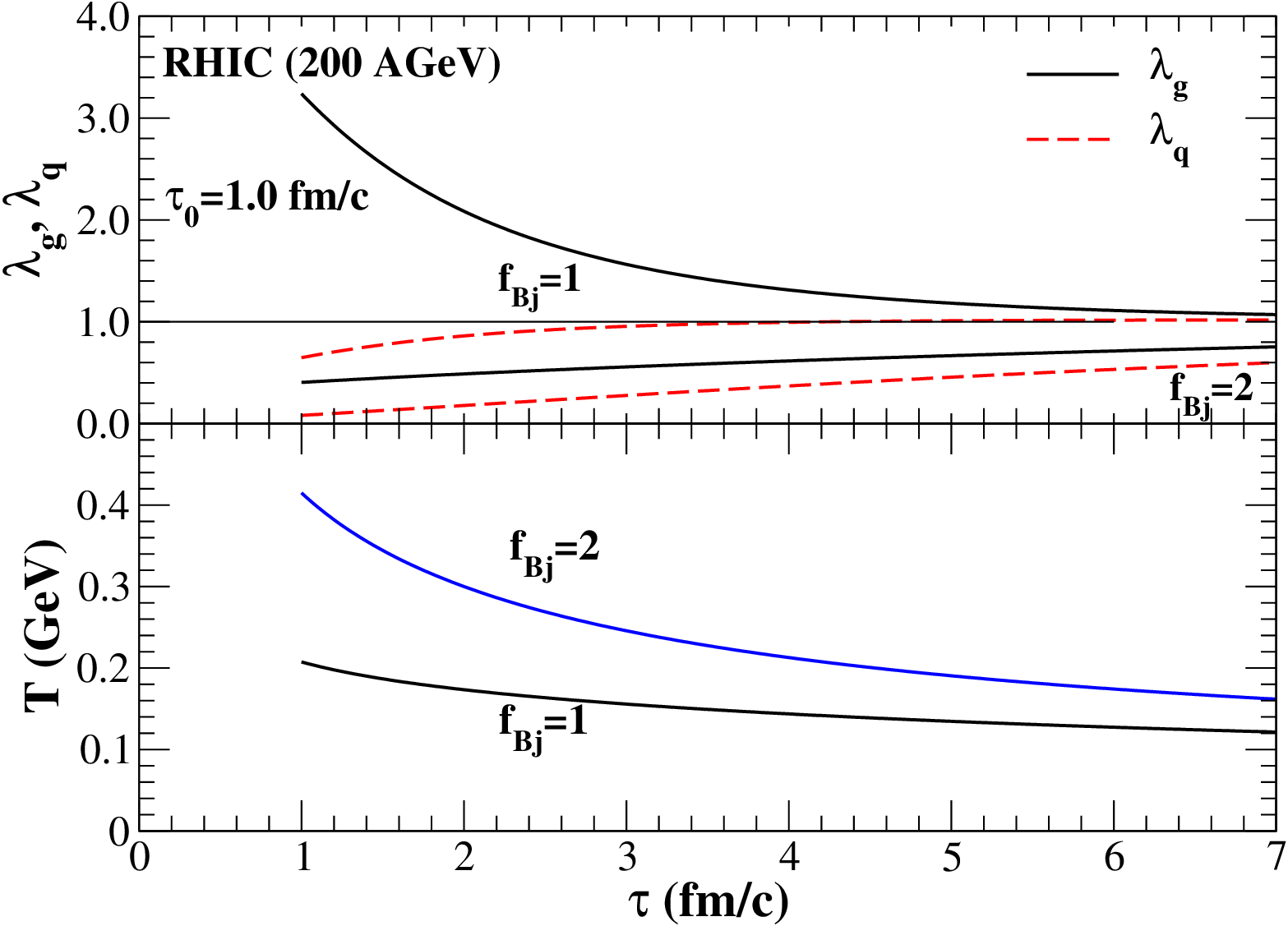}}
\centerline{\includegraphics*[width=8.0 cm]{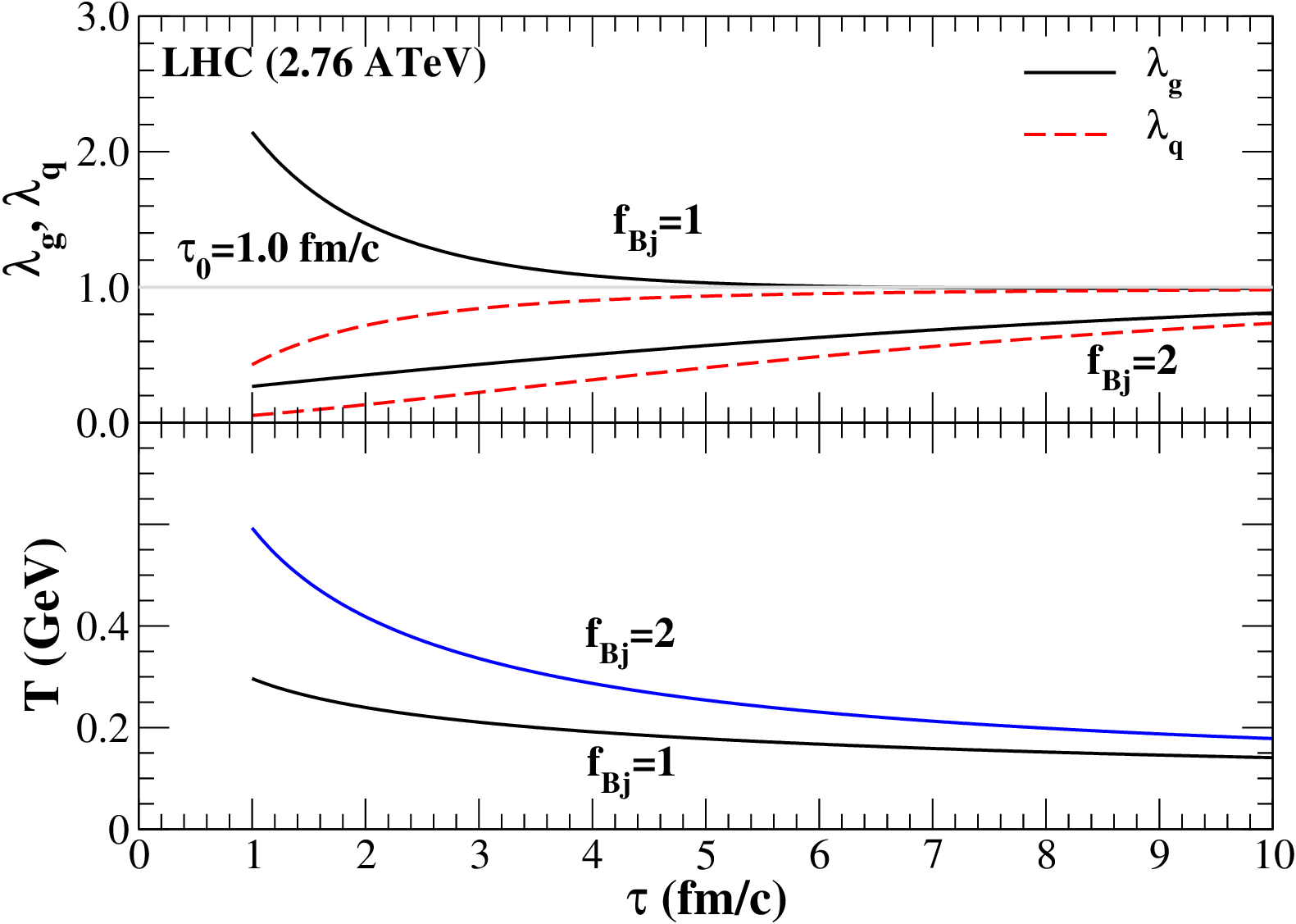}}
\centerline{\includegraphics*[width=8.0 cm]{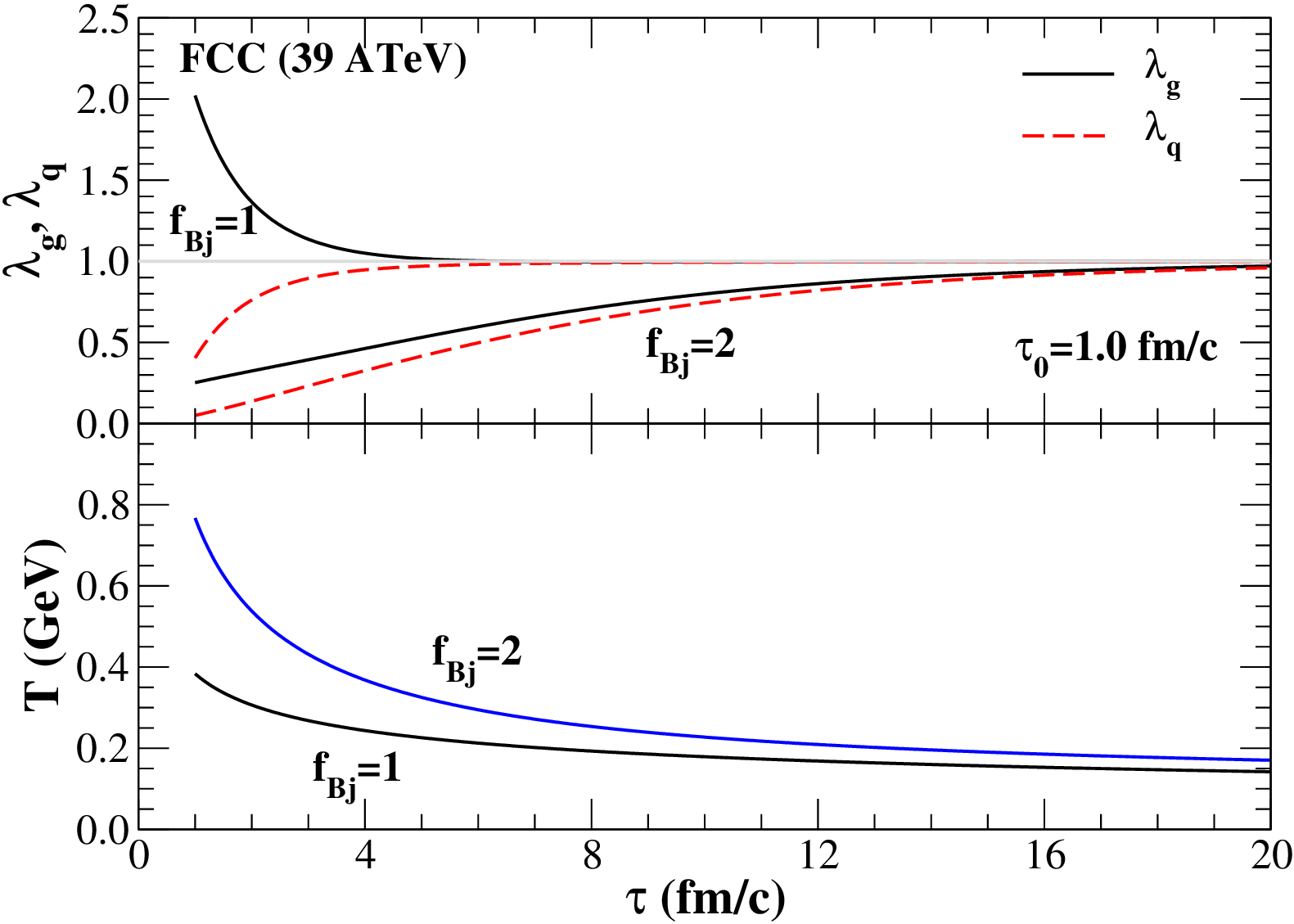}}
\caption{(Colour online) The time evolution of fugacities and temperature
at RHIC (200 AGeV), LHC (2.76 ATeV) and FCC (39 ATeV) 
with the assumption of formation time, $\tau_0=$ 1.0 fm/$c$ and 
the correction factor, $f_{\textrm {Bj}}=$1 and 2.  We have additionally assumed that
 $\lambda_g(\tau_0)=5\lambda_q (\tau_0)$} 
\label{fbj_lam_alpha_tau1}
\end{figure}
\begin{figure}
\centerline{\includegraphics*[width=8.0 cm]{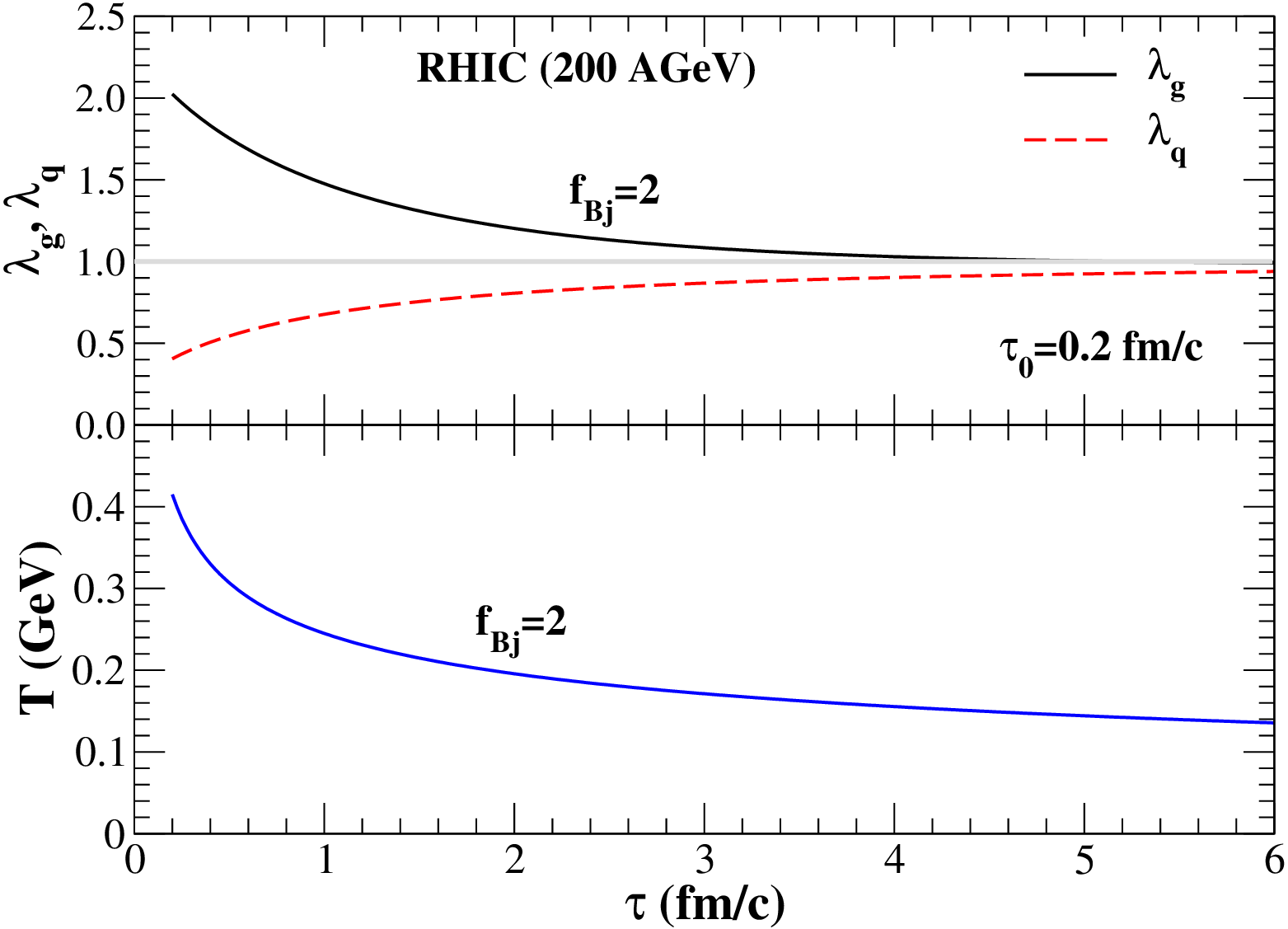}}
\centerline{\includegraphics*[width=8.0 cm]{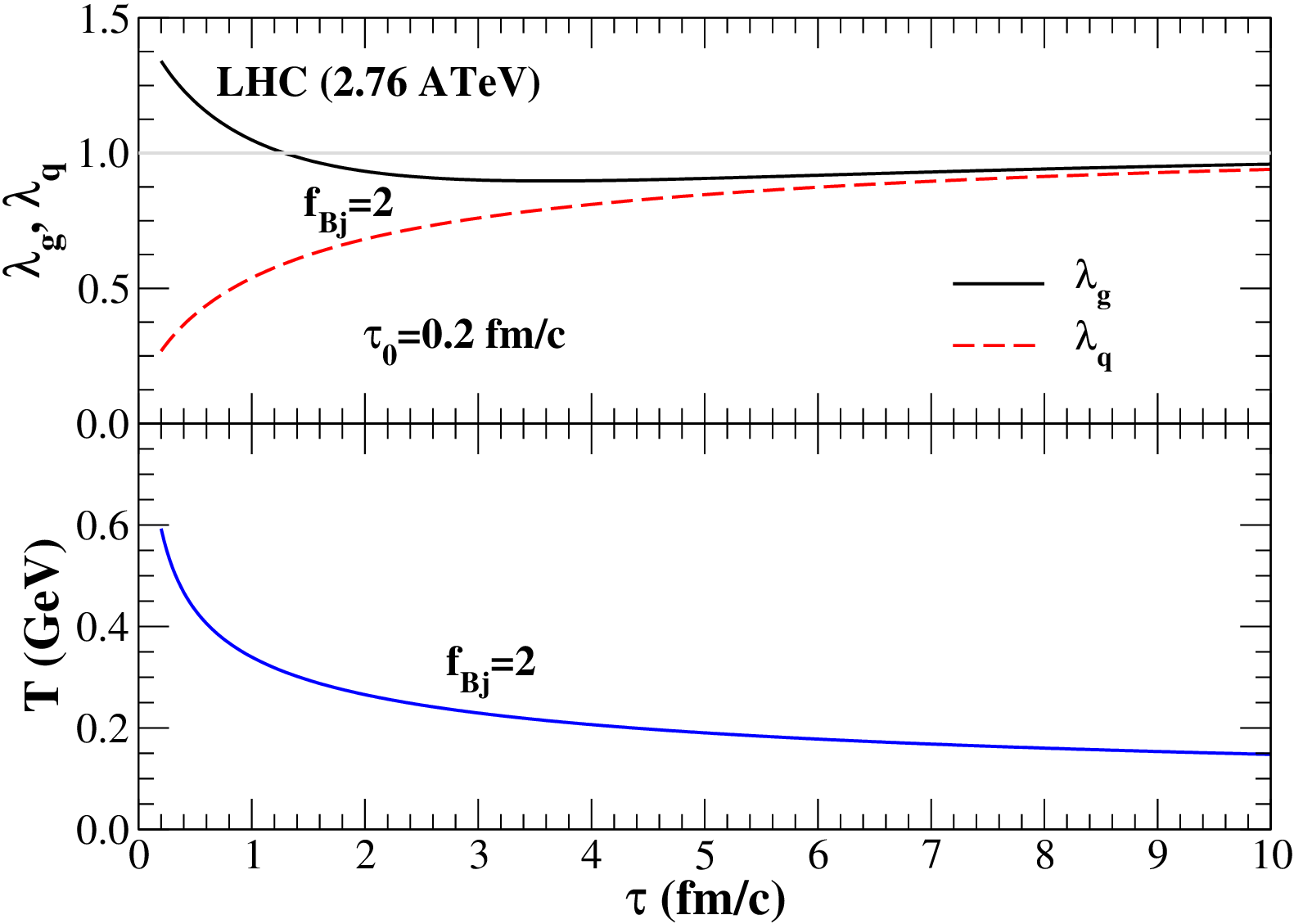}}
\centerline{\includegraphics*[width=8.0 cm]{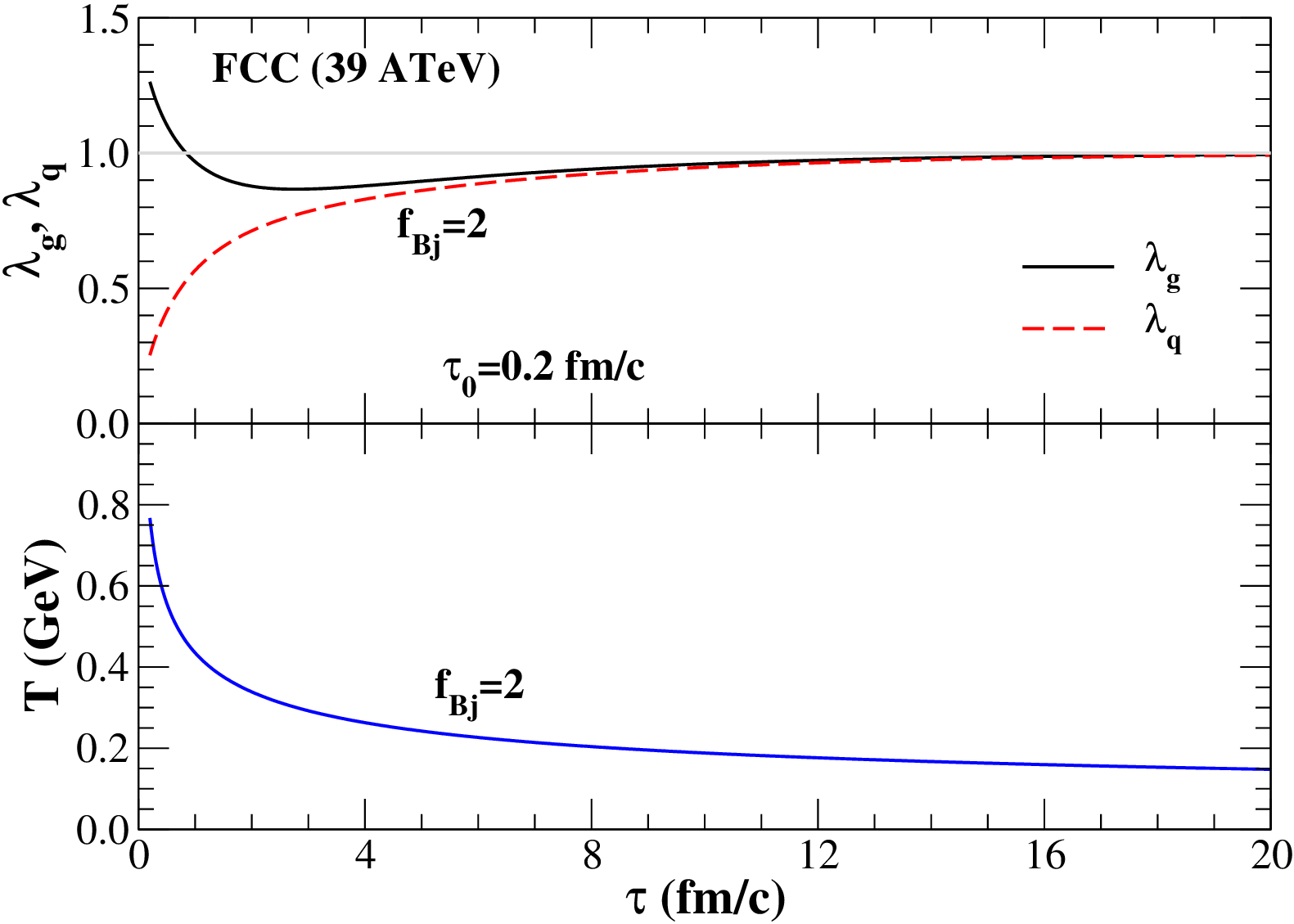}}
\caption{(Colour online) The time evolution of fugacities and temperature
at RHIC (200 AGeV), LHC (2.76 ATeV) and FCC (39 ATeV) with the
assumption of formation time, $\tau_0=$ 0.2 fm/$c$ and 
the correction factor, $f_{\textrm {Bj}}=$1 and 2. We have additionally assumed that
 $\lambda_g(\tau_0)=5\lambda_q (\tau_0)$ (see discussion)}
\label{fbj_lam_alpha_tau0.2}
\end{figure}
\begin{figure}
\centerline{\includegraphics*[width=8.0 cm]{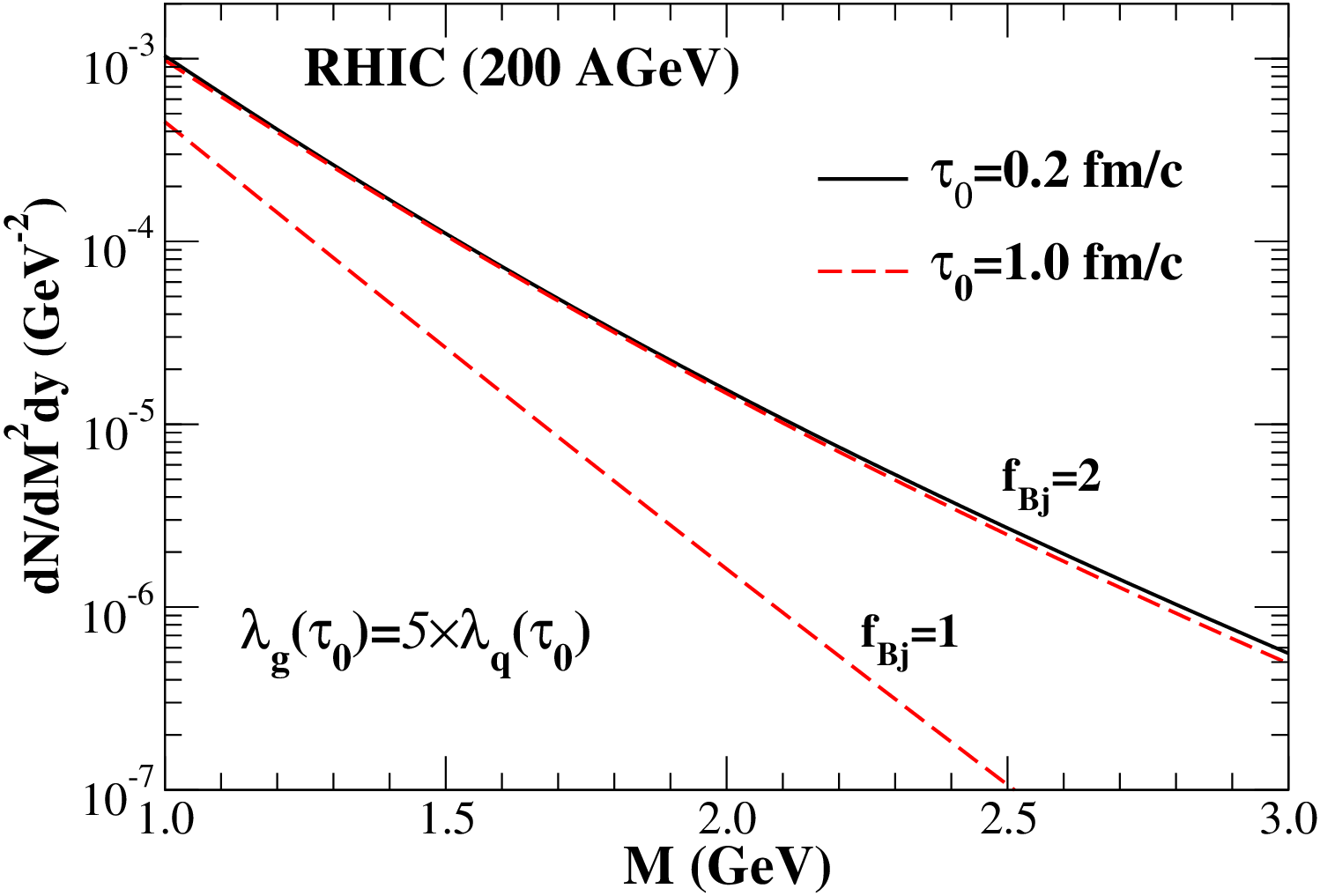}}
\centerline{\includegraphics*[width=8.0 cm]{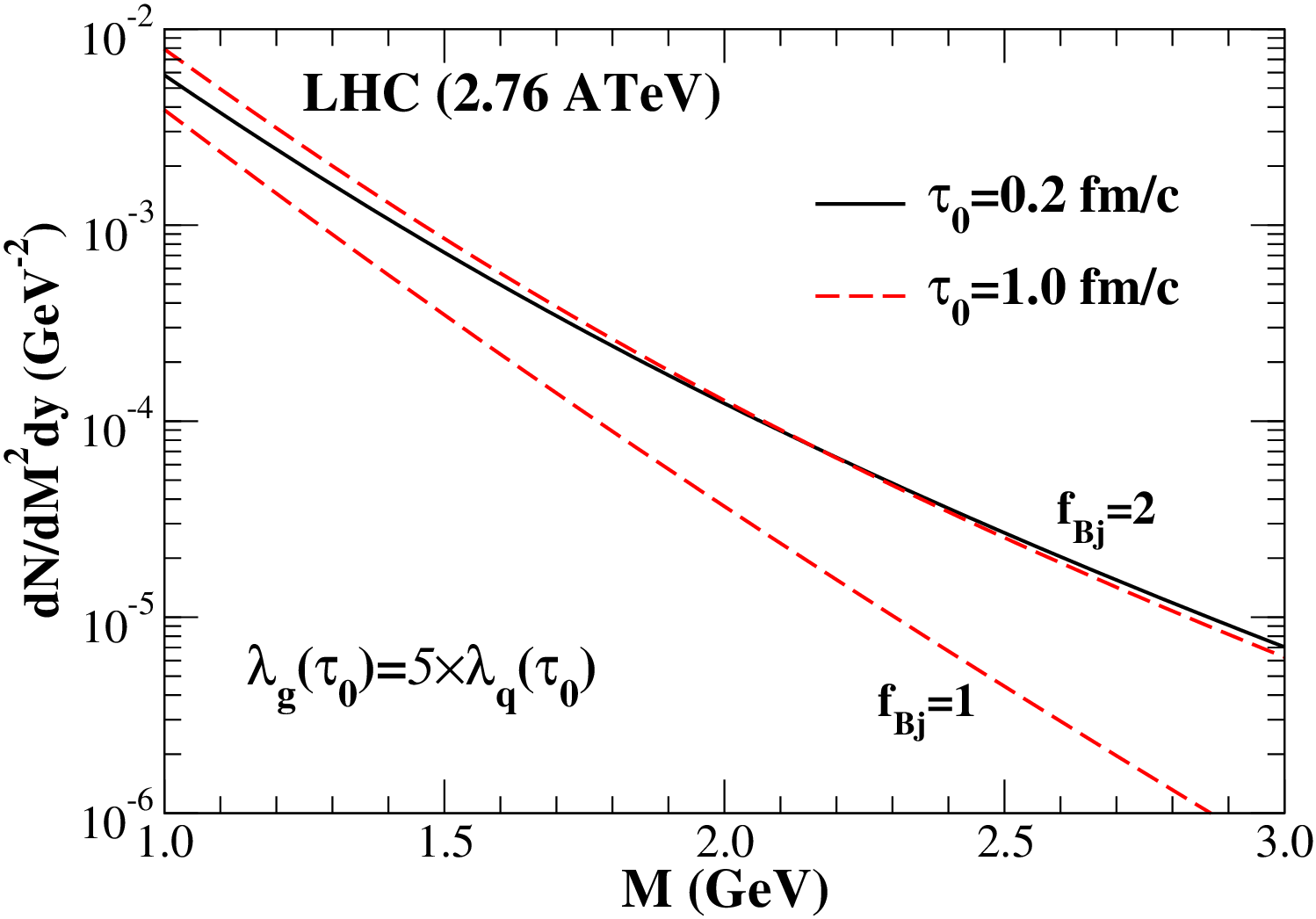}}
\centerline{\includegraphics*[width=8.0 cm]{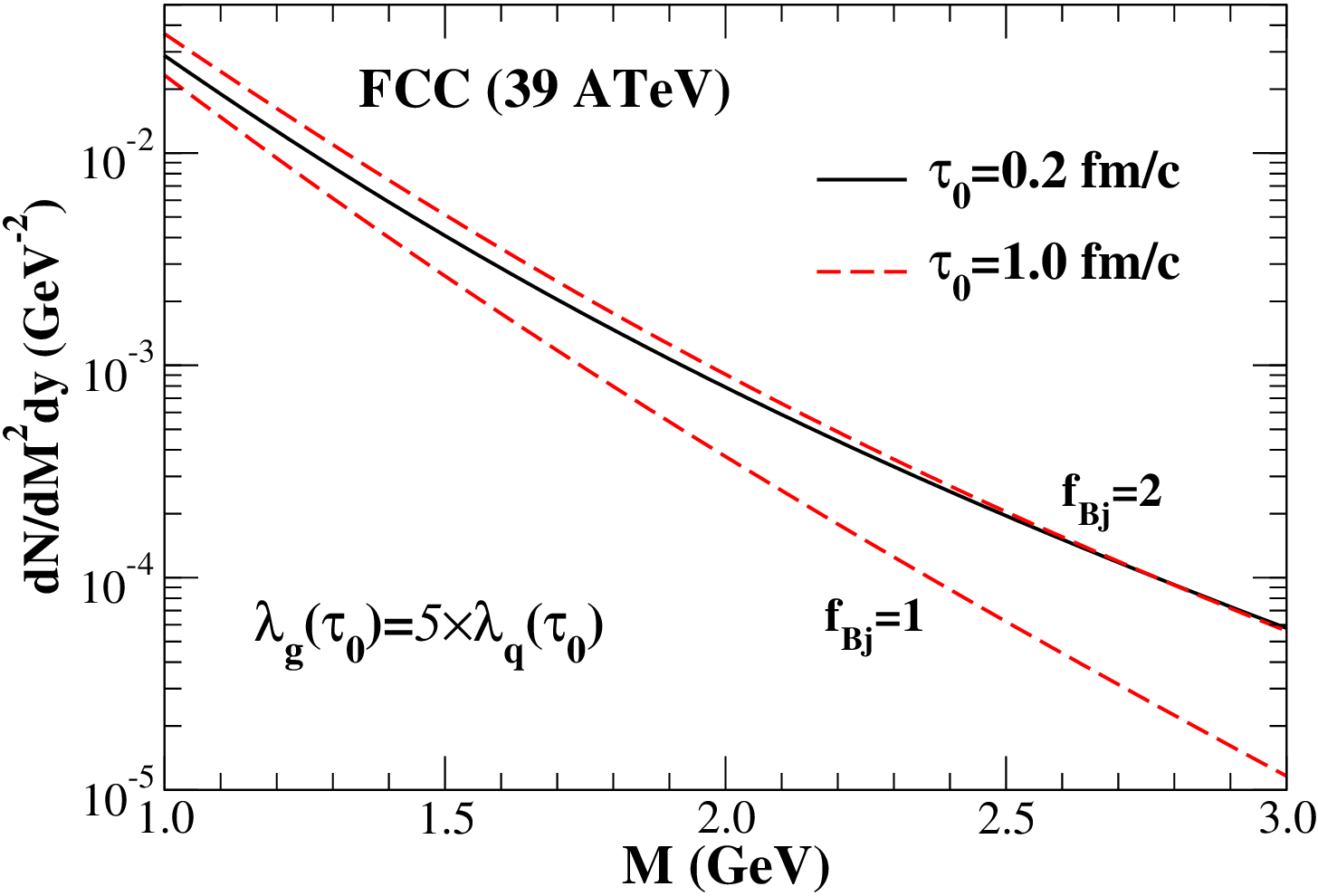}}
\caption{(Colour online) The invariant mass 
distribution of intermediate mass dileptons radiated from
nucleus-nucleus collision at RHIC (200 AGeV), LHC (2.76 ATeV) 
and FCC (39 ATeV) when of formation times
is reduced from 1 fm/$c$ to 0.2 fm/$c$ and the correction
factor $f_{\textrm {Bj}}$ increased from 1 to 2. The initial
fugacity of the gluons is assumed to be five times that for
the quarks.}
\label{dil_lam_alpha}
\end{figure}
In Figs.~\ref{fbj_lam_alpha_tau1} and ~\ref{fbj_lam_alpha_tau0.2} we show the time evolution of the fugacities and temperatures for RHIC (200 A GeV), LHC (2.76 ATeV) and FCC (39 ATeV) assuming $\tau_0$ to be 1 fm/$c$ and 0.2 fm/$c$ and the correction factor $f_{\textrm {Bj}}$ as 1 and 2 as before for the present case.

The results are very revealing and interesting to say the least. When $f_{\textrm {Bj}}$ is 
taken as 1 and the formation time is taken as 1 fm/$c$ gluons are over-saturated and quarks are under-saturated in the beginning. However, the $gg \leftrightarrow ggg$ and the $ gg \leftrightarrow q\overline{q}$ reactions included in our study quickly drive the plasma to near chemical equilibration at all the energies  under consideration. 

Increasing the $f_{\textrm {Bj}}$ to 2 produces the plasma in a state of under-saturation but again the system is driven to the chemical equilibrium at all the energies. The essential difference in the cases is that gluons remain  over-saturated in the early stages in the first case while they remain under-saturated
in the second case. 
the evolution.

The results get a lot more interesting and intriguing  when the formation time is assumed as
0.2 fm and $f_{\textrm {Bj}}$ remains 1. We have already seen above that the initial gluon and even quark fugacities remain considerably larger than unity for all the energies. This has a very amusing consequence. Now the  rate equations provide that the quark fugacities turn negative(!) for a part of the history
before becoming positive again! This is definitely unphysical, provided that our rate equations remain valid even under such extreme conditions.

Thus we show our results only for $f_{\textrm {Bj}}$ equal to 2. Now the  gluons are moderately over-saturated and quarks are moderately under-saturated during the initial stages and the partonic reactions quickly drive the system towards a state of chemical  equilibrium well before the plasma starts to hadronize.

An indication of this possibility was actually already present in the  earlier expression for $R_3$ (see Ref.\cite{biro}) as it contained a term $\sqrt{\lambda_g (2-\lambda_g)}$ because of some of the simplifying assumptions
made while deriving this. This would have driven the rate equation to  unphysical region for $\lambda_g >$ 2. In fact this originally prompted us to use the more recent estimate for $R_3$ (Eq.~\ref{R_3}) from Ref.~\cite{wang} which is free from this debilitating condition, as we encountered $\lambda_g >2$ very early in our
studies.  A far better and complete expression for the momentum distribution of the radiated gluons in the process $gg \leftrightarrow ggg$ is now available Ref.~\cite{ggg} and thus  a more accurate  expression for $R_3$ can be derived. We postpone this study for a future publication.

Finally we show our results for the production of intermediate
mass dileptons for the top RHIC energy, LHC (2.76 ATeV) and FCC (39 ATeV)
as before but only for those cases where the fugacities remain
in a physical domain during the evolution (see Fig.~\ref{dil_lam_alpha}).

\begin{table} 
\begin{center}
\begin{tabular}{|l|c|c|c|c|}
\hline
\multicolumn{5}{|c|} {$f_{\rm {Bj}}=1$} \\
\cline{1-5}
\hline
 Energy  & \multicolumn{2} {c |}{$\lambda_g(\tau_0)$= $\lambda_q(\tau_0)$} &  \multicolumn{2}{c|}{ $\lambda_g(\tau_0)$= 5$\lambda_q(\tau_0)$} \\
\cline{2-5}
(ATeV) & $T_0$ (GeV) & $\lambda_g(\tau_0)\tau_0$ (fm) & $T_0$ (GeV) & $\lambda_g(\tau_0)\tau_0$ (fm) \\
\hline
RHIC (0.2) & 0.196 & 2.046  & 0.207 & 3.239  \\
\hline
LHC (2.76) & 0.280  & 1.355 & 0.296 & 2.145 \\
\hline
LHC (5.02)  &  0.292  & 1.413  &  0.309 & 2.237   \\
\hline
 FCC (39)& 0.362 & 1.277 & 0.384  & 2.021  \\
\hline
\multicolumn{5}{|c|} {$f_{\rm {Bj}}=2$} \\
\cline{1-5}
\hline
 Energy  & \multicolumn{2} {c|}{$\lambda_g(\tau_0)$= $\lambda_q(\tau_0)$} &  \multicolumn{2}{c|}{ $\lambda_g(\tau_0)$= 5$\lambda_q(\tau_0)$} \\
\cline{2-5}
(ATeV) & $T_0$ (GeV) & $\lambda_g(\tau_0)\tau_0$ (fm) & $T_0$ (GeV) & $\lambda_g(\tau_0)\tau_0$ (fm) \\
\hline
RHIC (0.2) & 0.392 & 0.255 & 0.415 &  0.405 \\
\hline
LHC (2.76) & 0.560  & 0.169  & 0.593  & 0.268 \\
\hline
LHC (5.02)  & 0.585 & 0.177  &   0.619 & 0.280   \\
\hline
FCC (39)& 0.725 & 0.160 & 0.767 & 0.253  \\
\hline
\multicolumn{5}{|c|} {$f_{\rm {Bj}}=3$} \\
\cline{1-5}
\hline
 Energy  & \multicolumn{2} {c|}{$\lambda_g(\tau_0)$= $\lambda_q(\tau_0)$} &  \multicolumn{2}{c|}{ $\lambda_g(\tau_0)$= 5$\lambda_q(\tau_0)$} \\
\cline{2-5}
(ATeV) & $T_0$ (GeV) & $\lambda_g(\tau_0)\tau_0$ (fm) & $T_0$ (GeV) & $\lambda_g(\tau_0)\tau_0$ (fm) \\
\hline
RHIC (0.2) & 0.588  &  0.076 & 0.622 & 0.120 \\
\hline
LHC (2.76) &  0.840  & 0.050 & 0.889 &  0.079 \\
\hline
LHC (5.02)  & 0.877    & 0.052 & 0.929  & 0.083  \\
\hline
 FCC (39)& 1.087 & 0.047 & 1.150  & 0.075   \\
\hline
\end{tabular}


\caption{Initial temperature and product of gluon fugacity and formation time for different values of $f_{\rm {Bj}}$ (see text) at RHIC, LHC and FCC energies. }
\label{table1} 
\end{center}
\end{table}

We notice once again that the radiation of intermediate mass dileptons
is fairly robust against reasonable variations of formation time for a given $f_{\rm {Bj}}$
and the history of evolution of the fugacities. However it does 
depend strongly on $f_{\textrm {Bj}}$ principally as it immediately
gives rise to larger initial temperature.

\begin{table} 
\begin{center}
\begin{tabular}{|l|c | c | c  |c|}

\hline
 Energy (ATeV) & \hspace{0.1 cm} $\tau_0$ (fm) \hspace{0.1 cm} & \hspace{0.3cm} $f_{\rm {Bj}}$ \hspace{0.2cm} & \hspace{0.1 cm} $T_0$ (GeV) \hspace{0.1 cm} & \hspace{0.1 cm}  $\lambda_g=\lambda_q$ \hspace{0.1 cm}  \\

\hline
RHIC (0.2) & 0.2  &  2.82 & 0.552 & 0.457 \\
\hline
RHIC (0.2) & 0.6  &  2.05 & 0.401 & 0.398 \\
\hline
RHIC (0.2) & 1.0  &  1.88 & 0.369 & 0.307 \\
\hline
LHC (2.76) &  0.2  &  3.43  & 0.960 &  0.168 \\
\hline
LHC (2.76)  & 1.0    & 2.09 & 0.585  & 0.149  \\
\hline
 FCC (39) & 0.2 & 3.43 & 1.242  & 0.158   \\
\hline
 FCC (39) & 1.0 & 2.03 &  0.737  & 0.152   \\
\hline

\end{tabular}

\caption{Initial temperature and product of gluon fugacity and formation time for different values of $f_{\rm {Bj}}$ (see text) at RHIC, LHC and FCC energies. }
\label{table2} 
\end{center}
\end{table}

\section{Estimating $f_{\textrm {Bj}}$ from hydrodynamic calculations}

In the above we have reported detailed studies of the initial conditions, history of evolution,
and radiation of dileptons for several cases with some assumed vales for $f_\mathrm{Bj}$ etc.

We have seen explicitly that the
history of evolution of the system will be decided by the initial energy and  entropy 
density. This is of-course no surprise.  We have also seen that the assumption of thermal equilibration of
the plasma opens up a way for reasonable excursions in the description
of the evolution of the system, with reasonable assumptions about the
state of chemical equilibration. 
Thus we saw a very important result
which is not often clearly stated: once we assume that the plasma
is thermally equilibrated and can guess the relative initial fugacities of
quarks and gluons, ($\lambda_q/\lambda_g$), the initial temperature can be
uniquely fixed {\em provided} we additionally know $f_{\textrm {Bj}}$ or alternatively $\epsilon(\tau_0)$,
 for a given particle rapidity density. 
An independent guess about the formation time can then fully determine the
initial state.

The correction factor $f_{\rm {Bj}}$, needed to estimate the initial energy density should be necessarily more than 1, except when the life time of the system is very small or when the initial temperature is small. It has been argued~\cite{kari} that if transverse expansion can be ignored, one may estimate this correction factor as ($T_i/T_c$) where $T_c$  is the transition temperature of the plasma. In principal the transverse energy could decrease further till the system decouples, but it is felt that this could be compensated by transverse expansion.  We feel that if the life time  is sufficiently large or if the initial temperature is sufficiently large, the transverse expansion may set in well before the temperature drops to $T_c$, and thus the factor could be smaller than ($T_i/T_c$).

However in  view of these arguments the results for $f_{\rm {Bj}}$=1 in Table.~\ref{table1} , serve as a base-line for further discussion, at least for top RHIC and higher energies. In Table.~\ref{table1} we summarize values of the initial temperature $T_0$, and the product $\lambda_g(\tau_0)\tau_0$ for three typical values of  $f_{\rm {Bj}}$, viz 1, 2, and 3 and for $\alpha$ equal 1 and 0.2, as an easy reference. One can use it to use any desired value of $\tau_0$, for further studies.

We realize that a large value of $f_{\rm {Bj}}$, e.g., 2 or 3 necessarily leads to a large value for the initial temperature for the given multiplicities. Further, if we assume that the plasma is chemically equilibrated, at least for gluons, then the formation times are necessarily only a small fraction of a fm/c. A larger formation time would then also necessarily imply an under-saturated plasma. We feel that these aspects have not received sufficient attention in the literature. 

Looking closely we also realize this possibility should arise naturally, simply because a small formation time necessarily arises due to produced by vehement collisions among the partons and that should drive the system to chemical equilibration. The reverse would also be true.

\begin{figure}
\centerline{\includegraphics*[width=8.0 cm]{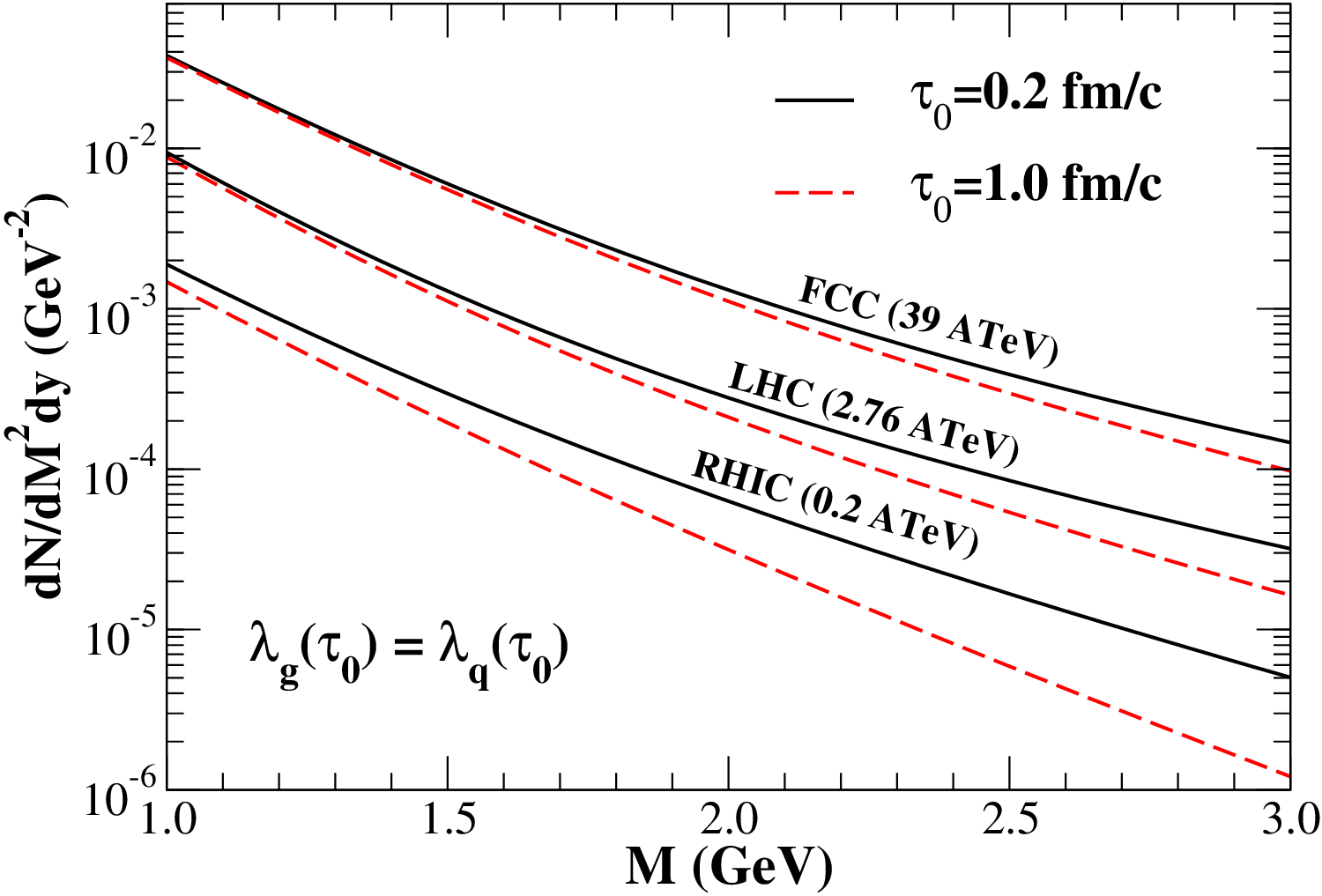}}
\caption{(Colour online) The invariant mass 
distribution of intermediate mass dileptons radiated from
nucleus-nucleus collision at RHIC (200 AGeV), LHC (2.76 ATeV) 
and FCC (39 ATeV) when of formation times
is reduced from 1 fm/$c$ to 0.2 fm/$c$.}
\label{dil_hydro}
\end{figure}
It may sound trivial, but not necessaryly so if we recall the
enormous developments in the hydrodynamic description of the relativistic 
heavy ion collisions. Thus for example in most of the highly successful
hydrodynamic studies the initial number of partons or their energy densities  are assumed to be produced~\cite{uli,cs}
according to a weighted sum of number of collisions and number of participants,
a fact which is independently and empirically established from the experimental
results on centrality dependence of particle rapidity density.
Thus, one writes:
\begin{equation}
s(\tau_0,x,y,b)=\kappa \, \left[ \,\alpha \, n_{\rm {BC}}(x,y,b) + (1-\alpha) \, n_{\rm {WN}}(x,y,b) \, \right]
\end{equation}
where, $\kappa$ is a normalization constant, $\alpha$ ($\sim$ 0.25 at top RHIC energy) denotes the fraction of hard contribution and $n_{\rm {BC}}$ and $n_{\rm {WN}}$
stand for number of binary collisions and number of wounded nucleons at the position $(x,y)$ for  impact parameter $b$.
The adjustable constant  $\kappa$ is then tuned to correctly reproduce the particle spectra and 
particle rapidity density for several classes of centrality. In some studies, alternatively energy density is similarly adjusted. 

These initial studies have been brought to a high degree of sophistication by inclusion
of the interesting phenomenon of  event by event fluctuations and shear as well as bulk viscosity. 
The hydrodynamic evolution only needs the initial energy density and pressure along with the equation of state.

In the following we attempt to get an estimate of $f_{\textrm {Bj}}$ by 
comparing the average energy density attained in these studies with
the Bj\"{o}rken energy density.

We obtained the average energy densities attained for RHIC and LHC energies by tuning the entropy density as discussed above  to reproduce the experimental multiplicity and spectra of the produce particles with assumption of different formation times (0.2, 0.6 and 1.0 fm/c)~\cite{cs}. We define:
\begin{equation}
\epsilon_0(\tau_0)= \frac{1}{\pi R_T^2} \int \epsilon(x,y) dx dy
\end{equation}
where $R_T$ is the transverse dimension. Comparing this energy density with the experimental Bj\"{o}rken energy density, we estimate the factor $f_{\rm {Bj}}$ (see Table~\ref{table2}).

 We note that for reasonable $\tau_0$ the $f_{\rm {Bj}} \ge $2.  We also note that  $f_{\rm {Bj}}$ is larger  for smaller $\tau_0$. This we feel happens   as the initial temperature and thus the pressure is large and the system losses more transverse energy in the work done against the pressure. We also find that the system necessarily produced in a chemically under-saturated state (note that we have assumed $\lambda_g(\tau_0)=\lambda_q(\tau_0)$).

We have used these initial conditions to solve the rate equations to get the history of evolution of the systems and in  Fig.~\ref{dil_hydro} we plot the dilepton spectra for all the cases under study as before. We see a smaller production for larger formation times. We also find that at least for RHIC and LHC energy the differences are substantial while larger differences are evident  only at larger invariant masses at FCC. 

Before closing this section we would like to emphasize another aspect of these considerations which needs more detailed elucidation. 

Almost all the studies for the electromagentic radiations assume that the energy density obtained following the procedure just discussed is for a chemically equilibrated plasma. This then leads to a value for the initial temperature $T_\mathrm{eq}$ obtained from the energy density $\epsilon_0(\tau_0)$. This temperature ($T_\mathrm{eq}$) would then be given by:
\begin{equation}
T_\mathrm{eq}^4 = \lambda_g T_0^4 \, .
\end{equation}

We immediately see that $T_\mathrm{eq}/T_0$ is equal to $\lambda_g^{1/4}$ 
 and thus $T_\mathrm{eq}$ is 20\% to 60\% smaller than $T_0$ for the results given in Table~\ref{table2}.

The corresponding entropy density, however has an even more interesting behaviour. For the "equivalent" system of
partons the entropy density is given by:
\begin{equation}
s_\mathrm{eq} \sim T_\mathrm{eq}^3 \, ,
\end{equation}
while for the actual system it is given by:
\begin{equation}
s_0 \sim \lambda_g T_0^3 \, ,
\end{equation}
and thus
\begin{equation}
\frac{s_\mathrm{eq}}{s_0}= \frac{1}{\lambda_g^{1/4}}.
\end{equation}
Thus we see that "chemically equilibrated" plasma has an entropy density which is
 20\% to 60\% larger (!) than what we actually needed to reproduce the muliplicity and no longer
consistent with it.

In addition to introducing an inherent inconsistency, we recall that the spectra of electromagnetic radiations
(e.g. dileptons) are proportional to $\lambda_q^2$ (here we have assumed $\lambda_q(\tau_0) =\lambda_g(\tau_0)$ and
calculated the evolution)
while their slope is determined by the temperature. This
could lead to interesting differences. Furthermore, the intgrated yield of electromagnetic radiation would be
$\propto \lambda_q^2 T^4$ which also would be quite different.

A possible way out of the above impasse could be to perform an iterative calculation by
changing the formation time till $\lambda_g$ turns out to be unity, while the particle spectra and
the multiplicity are still correctly reproduced. However this would be strictly valid only under 
the assumption of the formation of a thermally and chemically equilibrated quark gluon plasma at
the corresponding formation time.

\section{ Summary and Conclusions}
We have explored the limits of initial temperature, formation time and extent of chemical equilibration of quark-gluon plasma at RHIC, LHC, and FCC energies. The persistent result that we find points to the fact that as the initial energy density could be a factor of upto 2 (or more) larger than the Bj\"{o}rken estimate, the only possibility of having a chemically equilibrated plasma would require that the formation time is a small fraction of a fm/c. Conversely, if the formation time is large, the plasma is most likely produced in a chemically under saturated state. Next we have explored the chemical evolution of the systems and obtained the dilepton spectra, which are found to be sensitive to initial densities. And finally we have obtained an estimate of initial energy density  of several systems from hydrodynamic calculations which reproduce the experimental particle spectra and multiplicity. This provides that the initial energy densities could be a factor of 2-3 times larger than Bj\"{o}rken energy density  estimated experimentally. These studies then help us to find the initial temperature and draw interesting conclusions about the chemical equilibration and formation time.  We have also noted that invariant mass spectra of dileptons can prove to be a valuable probe to substantiate these findings.

In principle the study reported here could be improved in several ways, e.g., introduction of transverse expansion, consideration of shear and bulk viscosity, and event-by-event fluctuations. The incorporation of viscosity may reduce the estimates of initial temperature. The accounting of event-by-event fluctuations will throw open the possibility of locally varying $\lambda_g$ and $\lambda_q$ as well as  $\tau_0$ and $T_0$. It should also be of interest to extend these studies to non-central collisions and
solving the rate equations, check whether this also affects the elliptc flow for example for thermal photons and thermal
dileptons~\cite{rupa}.

We conclude from these studies that there is a need to understand the extent of chemical equilibration and formation time more accurately before we can draw reliable conclusions about electromagnetic radiations from the plasma.

\begin{acknowledgments} 
DKS gratefully acknowledges the grant of Raja Ramanna Fellowship by the Department of Atomic Energy, India.
We thank Steffen Bass, Raghunath Sahoo and Ajit Srivastava for helpful discussions.     
\end{acknowledgments}

\end{document}